\documentclass[transmag]{IEEEtran}
\usepackage{latexsym}
\usepackage{graphicx}
\usepackage{hyperref}
\usepackage{cite}
\usepackage{amsmath,amssymb,amsfonts,amsthm}
\usepackage{algorithmic}
\usepackage{physics}
\usepackage{textcomp}
\usepackage[T1]{fontenc}
\usepackage{caption}
\usepackage{subcaption}
\usepackage{epstopdf}
\usepackage{overpic}
\usepackage{array}
\usepackage{color}
\usepackage{url}
\usepackage{bm}
\usepackage{bbm}
\usepackage[none]{hyphenat}

\def\Htran{\mbox{\tiny $\mathrm{H}$}}
\def\Ttran{\mbox{\tiny $\mathrm{T}$}}
\def\CN{\mathcal{N}_{\mathbb{C}}}
\def\Real{\mathbb{R}}
\def\Complex{\mathbb{C}}

\def\Ex{\mathbb{E}}
\def\imagunit{\mathsf{j}} 
\def\ktx{\boldsymbol{\kappa}}
\def\krx{\vect{k}}

\newcommand{\vect}[1]{{\bf{#1}}}

\theoremstyle{plain}
\newtheorem{theorem}{Theorem}

\newtheorem{lemma}{Lemma}
\newtheorem{corollary}{Corollary}

\newtheorem{example}{Example}

\IEEEoverridecommandlockouts

\graphicspath{{./images/}}

\def\BibTeX{{\rm B\kern-.05em{\sc i\kern-.025em b}\kern-.08em T\kern-.1667em\lower.7ex\hbox{E}\kern-.125emX}}
\begin{document}

\title{Spatial Characterization of Electromagnetic Random Channels}

\author{Andrea Pizzo, Luca Sanguinetti, \IEEEmembership{Senior Member, IEEE}, Thomas L. Marzetta,
\IEEEmembership{Life Fellow, IEEE}
\thanks{Part of this work was presented at the Asilomar Conference on signals, Systems and Computers, Pacific Grove, CA, 2021 \cite{PizzoASILOMAR20}.}
\thanks{A. Pizzo was with University of Pisa and New York University. He is now with Universitat Pompeu Fabra, 08005 Barcelona, Spain (email: andrea.pizzo@upf.edu).}
\thanks{L. Sanguinetti is with University of Pisa, 56122 Pisa, Italy (email: luca.sanguinetti@unipi.it).}
\thanks{T. Marzetta is with New York University, 11201 New York, USA (email: tom.marzetta@nyu.edu).}}

\IEEEtitleabstractindextext{\begin{abstract}
The majority of stochastic channel models rely on the electromagnetic far-field assumption, which allows to decompose the channel in terms of plane waves. The far-field assumption breaks down in future applications that push towards the electromagnetic near-field region, such as those where the use of electromagnetically large antenna arrays is envisioned. Motivated by this consideration, we show how physical principles can be used to derive a plane-wave scalar channel model that is also valid in the reactive near-field region. Precisely, we show that narrowband wave propagation through a three-dimensional scattered medium can be generally modeled as a linear and space-variant system. We first review the physics principles that lead to a closed-form deterministic plane-wave representation of the channel impulse response. This serves as a basis for deriving a stochastic representation of the channel in terms of statistically independent Gaussian random coefficients for spatially stationary random propagation environments. The very desirable property of spatial stationarity can always be retained in the radiative near-field region by excluding reactive propagation mechanisms confined in close proximity to the source. Remarkably, the provided stochastic representation is directly connected to the Fourier spectral representation of a general stationary spatial random field.
\end{abstract}

}

\maketitle

\section{Introduction}

\IEEEPARstart{U}{nderstanding} the foundations of wireless communications systems requires accurate, yet tractable, channel models that reflect their main characteristics and properties. 
Their development is crucial to achieve a genuine fusion of electromagnetic theory with communication theory~\cite{TomBookReview}, which is the basis of the wave theory of information at the intersection of the two disciplines~\cite{FranceschettiBook}.

The physics of electromagnetism sets the boundary of what wireless communication systems are capable of~\cite{Migliore}. 
Physically meaningful channels are obtainable from the wave equation, whose solution yields an eigendecomposition of the channel in terms of plane waves or spherical waves~\cite{ChewBook}. 
Unlike models based on a spherical wave expansion~\cite{Gustafsson,Glazunov,Molisch}, channel models that are based on plane waves allow to treat radio wave propagation as a linear system by leveraging Fourier theory and without the recourse to special functions~\cite{MarzettaISIT}.

Plane-wave models are historically linked to the far-field (Fraunhofer) propagation regime wherein wavefronts are approximated as locally planar~\cite{BalanisBook}. This has been extensively used in the past wireless research at sub-$6$~GHz frequency bands~\cite{Sayeed2002,Veeravalli,PoonDoF,PoonCapacity}. 
However, as communications scale up in frequency entering the millimeter-wave and sub-terahertz frequency bands~\cite{Rangan,Rappaport,Lozano}, antenna arrays become electromagnetically large (compared to the wavelength).
The plane-wave assumption breaks down naturally in this regime, with potentially dramatic effects on system performance. For example, the incorporation of the \emph{wavefront curvature} in line-of-sight (LoS) channels offer spatial multiplexing capabilities -- similar to the ones of non-line-of-sight (NLoS) channels -- even for a single user scenario~\cite{Lozano}. Research in this direction is taking place under the names of holographic multiple-input-multiple-output (MIMO)~\cite{PizzoJSAC20,PizzoTWC21,DardariHolographic}, large intelligent surfaces~\cite{LIS}, and reconfigurable intelligent surfaces~\cite{RIS}. 

Based on the above discussion, there is a common belief in the wireless community that plane-wave models are inadequate to describe future wireless networks. 
The objective of this paper is to show that is not correct. In fact, classical physics teaches us that wave propagation can \emph{always} be formulated in terms of plane waves irrespective of the communication range (i.e., even in the near-field region) and under arbitrary propagation environments~\cite{ChewBook,PlaneWaveBook}.
This result builds upon Weyl's decomposition of a spherical wave into plane waves~\cite{Weyl,ChewBook} and scattering matrix theory~\cite{Saxon,Gerjuoy,Kerns1976,NietoWolf}. For simplicity, we focus on scalar electromagnetic fields, which physically correspond to acoustic propagation in general~\cite{MarzettaIT}. Generalization to vector electromagnetic channels would allow incorporating polarization~\cite{MarzettaNokia,MarzettaVector}.

\subsection{Contributions}

We start by uncovering the fundamentals of scalar wave propagation theory in \emph{deterministic} environments. 
These are typically modeled by using ray tracing tools or numerical electromagnetic solvers~\cite{MolischBook}. Both are not analytically tractable~\cite{MarzettaISIT}. Similarly to~\cite{PoonDoF,HanlenTWC}, we consider a continuous-space model that enables a functional viewpoint of MIMO channels; spatial sampling and discrete formulation tend to hide fundamental results, which are otherwise revealed by a continuous analysis~\cite{PizzoJSAC20}.
Our development breaks down the entire wave propagation problem into three parts: 
\begin{enumerate}
\item The transmission of (possibly) infinite number of plane waves by a source distribution; 
\item The reception of another (possibly) infinite number of plane waves at receiver, upon interaction with the environment; 
\item A linear scattering operator mapping the input spectrum onto the output spectrum of plane waves.
\end{enumerate}
The analysis shows that the electromagnetic channel can generally be modeled as a linear and space-variant system that is fully described by its six-dimensional spatial impulse response $h(\vect{r},\vect{s})$ at point $\vect{r}$ due to a unit impulse (point source) applied at point $\vect{s}$. This is obtained \emph{exactly} as a four-dimensional (4D) Fourier plane-wave representation that is function of the two horizontal wavenumber coordinates at source and receiver -- each one parametrizing every transmit and receive directions. In agreement with~\cite{FranceschettiLandau}, electromagnetic channels have only an apparent full informational structure, which is subjected to a lower dimensional representation. In the above representation, Fourier transforms at source and receiver provide a map between the spatial and the wavenumber (or angular) domains. The entire effect of the propagation environment is captured by an angular kernel describing the coupling between every pair of transmit and receive directions.
Compared to previous plane-wave representations available in the wireless literature~\cite{Sayeed2002,PoonDoF}, ours is applicable even in the reactive near-field region and embodies the lower dimensionality of electromagnetic channels.

Deterministic characterization of the angular kernel applies only to a specific environment. Instead, a \emph{stochastic} description represents an environmental class with common physical properties, with every ensemble describing propagation into hypothetically different environments. A stochastic channel encompasses a large-scale fading and a small-scale fading. This paper only considers the small-scale fading. Any large-scale fading model can be applied verbatim if the array size at both ends does not exceed the size of the local scattering neighborhood~\cite[Sec.~3.6]{LozanoBook}.
Our development builds upon the two following assumptions: \emph{complex Gaussian distribution} and \emph{wide-sense spatial stationarity}. These yield a Rayleigh fading model where $h(\vect{r},\vect{s})$ is a spatially stationary circularly symmetric complex Gaussian electromagnetic random field~\cite{MarzettaISIT,PizzoJSAC20}. Unlike~\cite{PizzoJSAC20}, this paper accounts for the presence of a radiating source.
We show that the stochastic angular kernel has jointly Gaussian entries that are statistically independent from one direction to another.
Altogether, we obtain a Fourier spectral representation of the electromagnetic random channel that exactly describes $h(\vect{r},\vect{s})$ only \emph{asymptotically}, i.e., as the normalized array size (compared to the wavelenght) grows to infinity.  


\subsection{Outline of the Paper}

The manuscript is organized as follows. In Section~\ref{sec:LoS_channel}, we provide a linear-system theoretic description of LoS propagation environments and derive the Fourier plane-wave representation of the channel impulse response.
This is extended in Section~\ref{sec:NLoS_channel} to deterministic NLoS environments under arbitrary conditions. 
Stochastic propagation environments are introduced in Section~\ref{sec:statistical_model}. Customization of the developed channel model to a prescribed environmental class is exemplified in Section~\ref{sec:phy_model_spectral}.
Final discussions and possible extensions of this paper are set forth in Section~\ref{sec:conclusions}.

\subsection{Notation}

We use upper (lower) case letters for spatial-frequency (spatial) entities. Blackboard bold letters denote integral operators. Boldfaced letters indicate vectors and matrices. The superscripts $^{\Ttran}$ and $^{\Htran}$ stand for transposition and hermitian. $\odot$ denotes the the Hadamard product. 
$\Real^n$ and $\Complex^n$ denote the $n$-dimensional Euclidean spaces of real- and complex-valued numbers, $\Re(\cdot)$ and $\Im(\cdot)$ denote real and imaginary parts, $|\cdot|$ denotes absolute value, $\lceil x \rceil$ denotes the least integer greater than or equal to $x$, $\delta(x)$ is the Dirac delta function, $\delta_n$ is the Kronecker delta function.
Calligraphic letters are used for sets. 
$m(\mathcal{X})$ denotes the Lebesgue measure, $\mathbbm{1}_{\mathcal{X}}(x)$ is the indicator function.
A general point $\vect{r} = x \hat{\vect{x}} + y \hat{\vect{y}} + z \hat{\vect{z}}$ in $\Real^3$ is described by its Cartesian coordinates $(x,y,z)$ with $\|\vect{r}\| = \sqrt{x^2 + y^2 + z^2}$ the Euclidean norm. $\nabla^2 = \frac{\partial^2}{\partial x^2} + \frac{\partial^2}{\partial y^2} + \frac{\partial^2}{\partial z^2}$ is the scalar Laplace operator. $\Ex\{\cdot\}$ denotes the expectation operator. The notation $n \sim \CN(0, \sigma^2)$ stands for a circularly-symmetric complex-Gaussian random variable with variance $\sigma^2$.

 
 \section{Line-of-sight propagation} \label{sec:LoS_channel}



\begin{figure} [t!]
        \centering
	\begin{overpic}[width=.8\columnwidth,tics=10]{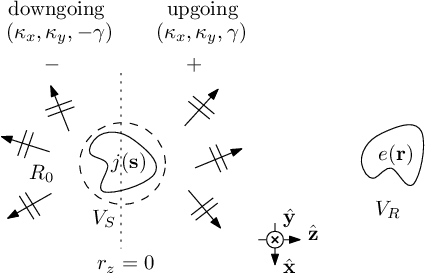}
\end{overpic} \vspace{-0.0cm}
                \caption{LoS wave propagation into a homogeneous and isotropic medium. A source density $j(\vect{r})$ creates an upgoing and a downgoing spectrum of plane waves traveling along the $z$- and $-z$-axis, respectively.}
                \label{fig:propagation_LoS} 
        \end{figure}

Consider a LoS propagation scenario wherein source and receiver are in visibility to each other due to the absence of any obstacle. This scenario is illustrated in Fig.~\ref{fig:propagation_LoS} and can be modeled as a linear and space-invariant (LSI) system, which is fully described by its channel impulse response at any point $\vect{r}$ \cite{MarzettaIT}. Linearity is due to the Maxwell's equations \cite{ChewBook}, while space-invariance is physically due to the fact that a LoS propagation environment appears the same irrespective of any space translation applied at source and/or receiver.


\subsection{Linear and Space-Invariant Electromagnetic Channels}

A space-time source density $j(\vect{r},t)$ occupies a physical volume ${V_S \subset \Real^3}$ and generates a scalar electric field $e(\vect{r},t)$. Propagation takes place into a three-dimensional (3D) homogeneous, isotropic and unbounded\footnote{For an unbounded medium we do not need to specify boundary conditions.} medium with velocity $c = 1/\sqrt{\mu \epsilon}$, $\mu$ and $\epsilon$ being the permeability and permittivity constants. 
The electric field must obey the inhomogeneous wave equation driven by $j(t,\vect{r})$. After taking a Fourier transform to both sides, this is equivalent to the inhomogeneous Helmholtz equation in the temporal-frequency domain: 
\begin{equation}  \label{Helm_eq}
\nabla^2 e(\omega,\vect{r}) + (\omega/c)^2 e(\omega,\vect{r}) = \imagunit \omega \mu j(\omega,\vect{r}).
\end{equation}
The above equation describes a linear and space-time invariant system \cite{MarzettaIT}, as the output spectrum is obtained by multiplying the input and output spectra
\begin{equation}  \label{Helm_eq_wavenumber}
E(\omega,\ktx) =  H(\omega,\ktx) \, J(\omega,\ktx)
\end{equation}
which is shown by taking a spatial Fourier transform\footnote{We use the convention $H(\omega,\ktx) =  \int_{\Real^3} \int_{-\infty}^\infty d\vect{r} dt \, a(\vect{r},t) e^{\imagunit (\omega t - \ktx^{\Ttran} \vect{r})}$ for space-time Fourier transforms. Time and space domains are mapped onto frequency and spatial frequency (or wavenumber) domains \cite[Sec.~1.2]{FranceschettiBook}.} of both sides of \eqref{Helm_eq}.
In \eqref{Helm_eq_wavenumber}, $H(\omega,\ktx)$ is the wavenumber-frequency response of the space-time invariant system given by
\begin{equation} \label{wave_freq_response} 
H(\omega,\ktx) = \frac{\imagunit \omega \mu}{\|\ktx\|^2 - (\omega/c)^2}.
\end{equation}
The essence of LoS propagation is so fully captured by \eqref{wave_freq_response}, which describes a two-poles system. Nevertheless, the majority of signals used in wireless communications are narrowband, implying that $J(\omega,\ktx)$ is a time-harmonic source at frequency $\omega$. Hence, we can neglect the temporal (frequency) dependence of \eqref{wave_freq_response}, which corresponds to a phasor notation in electromagnetism \cite[Sec.~1.1.3]{ChewBook}.

The resulting LSI system is completely characterized by its spatial response $h(\vect{r})$ due to an impulse (point source) applied at any point $\vect{r}$. This response must obey the inhomogeneous Helmholtz equation driven by $\delta(\vect{r})$:
\begin{equation}  \label{scalar_Helm_eq}
\nabla^2 h(\vect{r}) + \kappa^2 h(\vect{r}) = \imagunit \kappa \eta \delta(\vect{r}).
\end{equation}
where we introduced the wavenumber $\kappa=\omega/c$ and the wave impedance $\eta=\sqrt{\mu/\epsilon}$ constants. Here, $h(\vect{r})$ is found by solving the second-order partial differential equation in \eqref{scalar_Helm_eq}, whose solution is well-known and given by
\begin{equation} \label{impulse_response_LoS} 
h(\vect{r}) = -\imagunit \kappa \eta G(\vect{r})
\end{equation}
where $G(\vect{r})$ is the scalar Green's function \cite[Eq.~(1.3.42)]{ChewBook}
\begin{equation} \label{Green}
G(\vect{r}) = \frac{e^{\imagunit \kappa r}}{4 \pi r} = G(r)
\end{equation}
which only depends on $r = \| \vect{r}\|$. Physically, \eqref{Green} describes an \emph{outgoing scalar spherical wave} that propagates radially from the point source. In fact, the total phase shift accumulated by the spherical wave over a sphere of fixed radius $r$ is the same regardless of the observation point $\vect{r}$.
The input-output relationship between $e(\vect{r})$ and $j(\vect{r})$ is the spatial convolution
\begin{align}  \label{channel_LoS}
e(\vect{r}) &= \int_{\Real^3} d\vect{s} \, j(\vect{s}) h(\vect{r} - \vect{s}) 
\end{align}
where $h(\vect{r}-\vect{s})$ is the space-invariant channel impulse response in \eqref{impulse_response_LoS} at point $\vect{r}$ due to an impulsive input (point source) applied at $\vect{s}$.
Physically, the output field is described by an integral superposition of spherical waves each one of which is generated at point $\vect{s}$. Superimposing contributions generated by all points $\vect{s}\in V_S$ yields a \emph{non-planar wave} with some curvature. 

In summary, the field $e(\vect{r})$ created by any spatially distributed source $j(\vect{r})$ can be essentially described in terms of spherical waves.
We next show that the same argument is valid for plane waves.
  
\subsection{Impulse Response of LoS Electromagnetic Channels}
  
The Weyl's identity \cite{Weyl} relates a spherical wave to an uncountably infinite number of plane waves traveling to every directions \cite[Eq.~(2.2.27)]{ChewBook},
\begin{align} \label{Weyl_identity}
\frac{e^{\imagunit \kappa r}}{r}  &= \frac{\imagunit}{2\pi}  \iint_{-\infty}^{\infty} d\kappa_xd\kappa_y \, \frac{e^{\imagunit ( \kappa_x r_x + \kappa_y r_y + \gamma(\kappa_x,\kappa_y) |r_z|)}}{\gamma(\kappa_x,\kappa_y)}.
\end{align}
This is found by computing the inverse spatial Fourier transform of \eqref{Green} with respect to the $\kappa_z$-coordinate, now $\gamma(\kappa_x,\kappa_y)$, that is parametrized by the horizontal wavenumber coordinates $(\kappa_x,\kappa_y)\in\Real^2$ as 
\begin{equation} \label{kappa_z}
\gamma(\kappa_x,\kappa_y) = 
\begin{cases}
\sqrt{\kappa^2 - \kappa_x^2 - \kappa_y^2} &  \kappa_x^2 + \kappa_y^2\le \kappa^2 \\
\imagunit \sqrt{\kappa_x^2 + \kappa_y^2 - \kappa^2} &  \kappa_x^2 + \kappa_y^2> \kappa^2.
\end{cases}
\end{equation}
Notice that $\Re(\gamma) \ge 0$ and $\Im(\gamma) \ge 0$ in \eqref{kappa_z}, which are known as the Sommerfeld's radiation condition at infinity for an unbounded medium and ensure convergence of the improper integral in \eqref{Weyl_identity} \cite{ChewBook,PlaneWaveBook}.
The triplet $(\kappa_x,\kappa_y,\gamma)$ always satisfies the condition $\kappa_x^2+\kappa_y^2+\gamma^2=\kappa^2$ for all $(\kappa_x,\kappa_y)\in\Real^2$ and, after normalization, specifies the direction of propagation of each outgoing plane wave, namely 
\begin{align} \label{wavenumber_spherical}
\hat{\ktx}_{\pm} & = \hat{\vect{x}} \frac{\kappa_x}{\kappa}  + \hat{\vect{y}} \frac{\kappa_y}{\kappa}  \pm \hat{\vect{z}} \frac{\gamma(\kappa_x,\kappa_y)}{\kappa} \\& \label{wavenumber_spherical_2}
= \hat{\vect{x}} \sin\theta_{\rm t} \cos\phi_{\rm t} + \hat{\vect{y}} \sin\theta_{\rm t} \sin\phi_{\rm t}  + \hat{\vect{z}} \cos\theta_{\rm t}
\end{align}
where $(\theta_{\rm t},\phi_{\rm t}) \in [0,\pi] \times [0,2\pi)$ are the elevation and azimuth angles in the source reference frame. Due to the connection between the wavenumber and angular domains, we will refer to spatial frequencies or angles indistinctly.
For each direction, there are two types of plane waves, i.e., \emph{upgoing} and \emph{downgoing} plane waves. The former travel in the half-space $z>0$ (i.e., $\theta_{\rm t}\in[0,\pi/2]$) and are specified by a term $e^{\imagunit \gamma r_z}$. Instead, the latter travel in the half-space $z<0$ (i.e., $\theta_{\rm t}\in(\pi/2,\pi]$) and are of the form $e^{-\imagunit \gamma r_z}$. We will use the $+$ and $-$ convention to distinguish between quantities associated with upgoing and downgoing waves, respectively. 
The plane-wave decomposition of $h(\vect{r})$ is obtained by substituting the Weyl's identity \eqref{Weyl_identity} into \eqref{impulse_response_LoS} and is finally reported next.

\begin{lemma} \label{th:2d_plane_wave_representation_LoS}
The channel response $h(\vect{r})$ modeling a LoS propagation environment is exactly given by the 2D Fourier plane-wave representation 
\begin{align} \label{impulse_response_LoS_plane_Wave}
h(\vect{r})  & =  \frac{\kappa \eta}{2}  \iint_{-\infty}^{\infty} \frac{d\kappa_x}{2\pi}\frac{d\kappa_y}{2\pi}\, \frac{e^{\imagunit \left( \kappa_x r_x + \kappa_y r_y + \gamma(\kappa_x,\kappa_y) |r_z| \right)}}{\gamma(\kappa_x,\kappa_y)}
\end{align}
where $\gamma(\kappa_x,\kappa_y)$ is defined in \eqref{kappa_z}.
\end{lemma}

The channel impulse response is obtained as an integral superposition of upgoing and downgoing plane waves each one having angle-dependent amplitude $1/\gamma$. Notice that this representation is perfectly consistent with physics as plane waves are natural eigen-solutions of the Helmholtz equation \cite{ChewBook}. In the physics literature, representations in the form of \eqref{impulse_response_LoS_plane_Wave} are known as {angular representations} for obvious reasons \cite{Wolf_1959,Sherman,Devaney}. 
A linear system-theoretic interpretation of \eqref{impulse_response_LoS_plane_Wave} is given next. 

\subsection{Wavenumber Response and Migration Filter} \label{sec:wavenumber_response_LOS}

For any fixed $r_z \in \Real$, each plane wave can be regarded as a phase-shifted version of a 2D spatial-frequency Fourier harmonic, namely
\begin{align} \label{2D-plane-wave}
e^{\imagunit \left( \kappa_x r_x + \kappa_y r_y + \gamma |r_z| \right)}  = e^{\imagunit \left( \kappa_x r_x + \kappa_y r_y\right)}  e^{\imagunit \gamma |r_z|}
\end{align}
where the phase-shift is applied along the $z$-axis. 
Based on this observation, $h(\vect{r})$ in   \eqref{impulse_response_LoS_plane_Wave} can be rewritten in terms of its wavenumber response via a 2D inverse spatial Fourier transform 
\begin{equation} \label{impulse_response_Fourier}
h(\vect{r}) = \iint_{-\infty}^{\infty} \frac{d\kappa_x}{2\pi} \frac{d\kappa_y}{2\pi}  \, H(\kappa_x,\kappa_y) e^{\imagunit \left(\kappa_x r_x +  \kappa_y r_y \right)} 
\end{equation}
with spectrum
\begin{equation} \label{wavenumber_response_LoS}
H(\kappa_x,\kappa_y) = \frac{\kappa \eta}{2} \frac{1}{\gamma(\kappa_x,\kappa_y)} e^{\imagunit \gamma(\kappa_x,\kappa_y) |r_z|}
\end{equation}
where the dependence of the channel's spectrum on the parameters $r_z$ is omitted.
Due to the multiplicative nature of \eqref{wavenumber_response_LoS}, the LoS channel can be regarded as a cascade of two LSI systems having wavenumber responses given by $1/\gamma$ and $e^{\imagunit \gamma |r_z|}$. 
The former uniquely describes the channel at the $r_z=0$ plane and is due to the Helmholtz equation in \eqref{scalar_Helm_eq} that enforces a `bowl-shaped' behavior in the channel's spectrum. This agrees with the plane-wave nature of the channel for which the spherical constraint $\kappa_x^2+\kappa_y^2+\gamma^2 = \kappa^2$ holds. When parametrized on the $\kappa_x \kappa_y$-plane, $1/\gamma$ accounts for the area change of the parametrized surface element \cite{PizzoJSAC20}. 
At any non-zero $r_z$, the channel is obtained by passing $h(r_x,r_y,0)$ through an LSI system with wavenumber response $e^{\imagunit \gamma |r_z|}$. This filtering operation is known in physics as migration and the associated system as \emph{migration filter} \cite{PizzoTSP21}.  
This behaves as either an all-pass filter that simply introduces a phase shift or as a low-pass filter that cuts out spatial frequencies above certain values. This is because  $(\kappa_x,\kappa_y)$ can vary independently in $\Real^2$ and hence $\gamma$ in \eqref{kappa_z} is either real- or imaginary-valued. 
In particular, $\gamma$ is real-valued within
\begin{equation}  \label{disk_T}
\mathcal{D} = \{ (\kappa_x,\kappa_y)\in\Real^2 : \kappa_x^2 + \kappa_y^2 \le \kappa^2\}
\end{equation}
given by a disk of radius $\kappa = 2\pi/\lambda$ and imaginary-valued elsewhere. 
Hence, the exponential 
\begin{equation} \label{exponential}
e^{\imagunit ( \kappa_x r_x + \kappa_y r_y + \gamma |r_z|)} = e^{\imagunit ( \kappa_x r_x + \kappa_y r_y)}  e^{\imagunit \Re(\gamma) |r_z|} e^{- \Im(\gamma) |r_z|}
\end{equation}
 is either an oscillatory or an exponentially-decaying function in $r_z$ with decay factor proportional to $r_z/\lambda$. Plane waves with horizontal wavenumber coordinates $(\kappa_x,\kappa_y)\in\mathcal{D}$ are called \emph{propagating} -- due to their capability of propagating wirelessly over longer distances -- or \emph{evanescent} otherwise. 
 Hence, for communication ranges of at least a few wavelengths, the LoS channel begin showing a low-pass filtering behavior; see \cite[Fig.~1]{PizzoTSP21}. Moreover, since $H(\kappa_x,\kappa_y)$ depends only on $\kappa_x^2 + \kappa_y^2$ through $\gamma$ in \eqref{kappa_z}, this filter is of circular low-pass type, as summarized next.
 
 \begin{corollary} \label{th:bandlimited}
When evanescent waves are discarded, the LoS channel impulse response $h(\vect{r})$ is \emph{circularly bandlimited} with wavenumber bandwidth
\begin{equation} \label{circular_bandwidth}
m(\mathcal{D}) = \pi \kappa^2 = \frac{4 \pi^3}{\lambda^2}
\end{equation}
inversely proportional to the wavelength squared.
 \end{corollary}

An application of the above result is the generalization of the sampling theorem for bandlimited time-domain signals to spatial electromagnetic channels~\cite{PizzoTSP21}.

 \subsection{Fraunhofer Far-field Approximation}

        
 We now recall the conditions under which $h(\vect{r})$ becomes the Fraunhofer far-field model \cite{ChewFarField}.
 We consider a reference point $\vect{r}_0 = \hat{\vect{x}} r_{x,0} + \hat{\vect{y}} r_{y,0} + \hat{\vect{z}} r_{z,0}$ of radius $r_0=\|\vect{r}_0\|$ and observe $h(\vect{r})$ in a neighbourhood of this point, i.e., at all points $\vect{r} = \vect{r}_0-\vect{r}^\prime$ with $\vect{r}^\prime = \hat{\vect{x}} r_{x}^\prime + \hat{\vect{y}} r_{y}^\prime + \hat{\vect{z}} r_{z}^\prime$. 
The Weyl identity in \eqref{Weyl_identity} yields
\begin{align} \notag
& \frac{e^{\imagunit \kappa \|\vect{r}_0-\vect{r}^\prime\|}}{\|\vect{r}_0-\vect{r}^\prime\|}  =  \frac{\imagunit}{2\pi}  \iint_{-\infty}^{\infty} d\kappa_xd\kappa_y \, \frac{1}{\gamma(\kappa_x,\kappa_y)} \\& \label{far_field_1} \hspace{1cm} e^{\imagunit ( \kappa_x (x_0-x^\prime) + \kappa_y (y_0-y^\prime) + \gamma(\kappa_x,\kappa_y) |z_0-z^\prime|)}
\end{align}
given $\gamma(\kappa_x,\kappa_y)$ as in \eqref{kappa_z}. In the half-space $z^\prime>z_0$, 
\begin{equation} \label{far_field_2}
h(\vect{r}) =   \iint_{-\infty}^{\infty} d\kappa_xd\kappa_y \, \frac{F(\kappa_x,\kappa_y)}{\gamma(\kappa_x,\kappa_y)}  e^{\imagunit \kappa r_0 g(\kappa_x,\kappa_y)}
\end{equation}
where $F(\kappa_x,\kappa_y) = \frac{\kappa \eta}{2(2\pi)^2} e^{-\imagunit (\kappa_x x^\prime + \kappa_y y^\prime + \gamma(\kappa_x,\kappa_y) z^\prime)}$ and
\begin{equation} \label{h_fun}
g(\kappa_x,\kappa_y) = \frac{\kappa_x}{\kappa} \frac{x_0}{r_0}+ \frac{\kappa_y}{\kappa} \frac{y_0}{r_0} + \frac{\gamma(\kappa_x,\kappa_y)}{\kappa} \frac{z_0}{r_0}.
\end{equation}
When at least one of the Cartesian coordinates of $\vect{r}_0$ is much larger than the wavelength, i.e., $r_0/\lambda$ is very large, the term $e^{\imagunit \kappa r_0 g(\kappa_x,\kappa_y)}$ in \eqref{far_field_2} rapidly oscillates as a function of $(\kappa_x,\kappa_y)$ thus creating an almost zero net contribution due to several periodic cycles adding together destructively. Most of the contributions to the integral will come from around the stationary point of $g(\kappa_x,\kappa_y)$ where the function slowly varies \cite[Sec.~2.5.1]{ChewBook}.
Hence, the integral is amenable to the \emph{stationary phase} approximation based on which the only non-negligible contribution to \eqref{far_field_2} is around the stationary points of $g(\kappa_x,\kappa_y)$ where its partial derivatives are zero,
\begin{equation} \label{stationary_point}
(\kappa_{x,0},\kappa_{y,0}) = \left(\kappa \frac{x_0}{r_0},\kappa \frac{y_0}{r_0}\right)
\end{equation}
so that $\kappa_{z,0} = \kappa \, {z_0}/{r_0}$.
The stationary phase point describes a propagation direction $\hat{\ktx}_0 = \hat{\vect{x}} \, {x_0}/{r_0} + \hat{\vect{y}} \, {y_0}/{r_0} + \hat{\vect{z}} \, {z_0}/{r_0}$ that points toward the reference point $\vect{r}_0$. Hence, even though a source emanates plane waves in all directions, several wavelengths away from the source, only one (or a few at most) plane waves around the stationary point are important \cite{ChewFarField}. The contribution to the channel field from all other directions becomes negligible.
Pulling out the slowly varying part in \eqref{far_field_2} sampled at \eqref{stationary_point}, i.e., $F(\kappa_{x,0},\kappa_{y,0})$, and applying the Weyl identity \eqref{Weyl_identity} to the remaining integral yields
\begin{align} \label{far_field_approx}
h(\vect{r}) & \approx
-\imagunit \kappa \eta \frac{e^{\imagunit \kappa r_0}}{4 \pi r_0} e^{-\imagunit \frac{\kappa}{r_0} (x_0 x^\prime + y_0 y^\prime + z_0 z^\prime)} \\&  
=  \frac{-\imagunit \kappa \eta}{4 \pi r_0} e^{\imagunit \kappa (r_0 - \vect{r}^\prime \cdot \hat{\vect{r}}_0)}.
\end{align}
Thus, in the Fraunhofer far-field region, all points in a neighborhood of the reference point see plane waves coming from the same direction $\hat{\ktx}_0$. The level of this approximation depends on $r_0$. The larger the distance, the higher the approximation accuracy. As an example, for a squared observation region of size $L$~m, a maximum phase error of $\pi/8$ across this region requires $r_0 \ge 2 L^2/\lambda$ where the minimum value for which this approximation is valid is known as \emph{Fraunhofer distance} \cite[Eq.~(4.47)]{BalanisBook}.

Next, we show that the wave propagation problem can be modeled \emph{exactly} in terms of plane waves regardless the communication range and under arbitrary propagation conditions.

  
\section{Non line-of-sight propagation} \label{sec:NLoS_channel}
 
 \begin{figure*} [t!]
        \centering
	\begin{overpic}[width=.7\linewidth,tics=10]{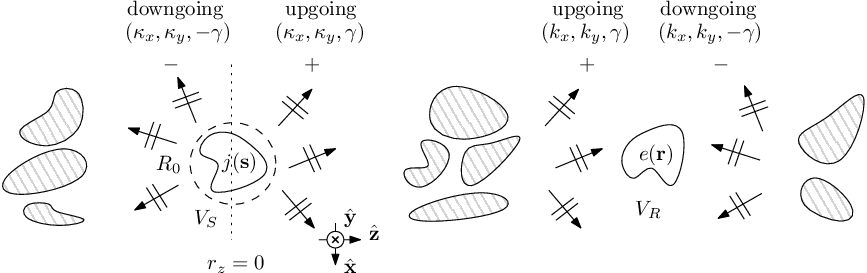}
\end{overpic} \vspace{-0.0cm}
                \caption{Wave propagation into a homogeneous and isotropic medium with scatterers (possibly) located everywhere in space.}
                \label{fig:propagation} 
        \end{figure*}

Wave propagation between communicating devices generally takes place in a NLoS environment due to the presence of scatterers in their surroundings. The transmitted plane-wave spectrum created by the source interacts with these objects through different propagation mechanisms (e.g., scattering, reflection, diffraction) before reaching the receiver distorted. We will refer to the interacting objects simply as scatterers, without distinguishing between the different types of interaction. The scatterers can be of any shape and size. The wave propagation problem can be studied \emph{exactly} in terms of plane waves and decomposed into three subproblems: \emph{i}) the creation of a transmit spectrum of plane waves by the source, \emph{ii}) the measurement of another receive spectrum of plane waves by the receiver, and \emph{iii}) the linear mapping between the two spectra~\cite{MarzettaNokia}.

Next, we elaborate on each subproblem separately and put them together to obtain a linear and space-variant (LSV) description of the channel.

\subsection{Fourier Plane-Wave Representation of Transmitted Field}

We evaluate the \emph{transmitted field} by the source at an intermediate point $\vect{r}^\prime = \hat{\vect{x}} r_x^\prime +  \hat{\vect{y}} r_y^\prime +  \hat{\vect{z}} r_z^\prime$ placed before any interaction with the scatterers could possibly occurs. Plugging \eqref{impulse_response_LoS_plane_Wave} at $\vect{r}^\prime$ into \eqref{channel_LoS}, we obtain
\begin{align}  \notag
e_{\rm t}(\vect{r}^\prime) & 
= \frac{\kappa \eta}{2 (2\pi)^2}
\iint_{-\infty}^{\infty} d\kappa_xd\kappa_y \,  \frac{e^{\imagunit \left( \kappa_x r_x^\prime + \kappa_y r_y^\prime\right)}}{\gamma(\kappa_x,\kappa_y)}  \\& \hspace{.5cm} \label{channel_LoS_2} \int_{\Real^3} d\vect{s} \, j(\vect{s})  e^{-\imagunit \left( \kappa_x s_x + \kappa_y s_y\right)} e^{\imagunit \gamma(\kappa_x,\kappa_y) |r_z^\prime-s_z|} .
\end{align}
Since source and receiver shall never be physically overlapped, outside of a sphere of radius $R_0 >0$ embedding $V_S$,  \begin{align}  \notag
& e_{\rm t}(\vect{r}^\prime) =  \\&\label{incident_field}
\begin{cases}
\iint_{-\infty}^{\infty} \frac{d\kappa_x}{2\pi}\frac{d\kappa_y}{2\pi} \,  E_{\rm t}^-(\kappa_x,\kappa_y)    e^{\imagunit \left( \kappa_x r_x^\prime + \kappa_y r_y^\prime -  \gamma r_z^\prime\right)}
\quad r_z^\prime < -R_0\\
\iint_{-\infty}^{\infty} \frac{d\kappa_x}{2\pi}\frac{d\kappa_y}{2\pi} \,  E_{\rm t}^+(\kappa_x,\kappa_y)   e^{\imagunit \left( \kappa_x r_x^\prime + \kappa_y r_y^\prime +  \gamma r_z^\prime \right)}
\quad r_z^\prime >  R_0
\end{cases}
\end{align}
where each plane wave has complex-valued amplitude
\begin{equation} \label{incident_spectrum}
E_{\rm t}^\pm(\kappa_x,\kappa_y) =  \frac{\kappa \eta}{2} \frac{J_\pm(\kappa_x,\kappa_y)}{\gamma}  
\end{equation}
with $J_\pm(\kappa_x,\kappa_y)$ the wavenumber spectrum of $j(\vect{r})$ obtained via a 3D spatial Fourier transform evaluated at $\kappa_z =  \pm \gamma$, i.e.,
\begin{align} \label{source_spectrum}
J_\pm(\kappa_x,\kappa_y) & = \iiint_{V_S} j(\vect{s}) e^{-\imagunit \left(\kappa_x s_x + \kappa_y s_y \pm \gamma s_z\right)}  d\vect{s}.
\end{align}
As a check, for a unit impulse (point source) located at the origin, i.e., $j(\vect{s}) = \delta(\vect{s})$, the use of the Weyl identity \eqref{Weyl_identity} into \eqref{incident_field} correctly yields the spherical wave solution in \eqref{impulse_response_LoS}.
In brief, the external effect of any current density is the creation of an outgoing spectrum of plane waves (propagating and evanescent). The corresponding transmitted field is given by the 2D \emph{Fourier plane-wave representation} in \eqref{incident_field} with a possible interpretation as either a plane-wave representation or an inverse 2D spatial Fourier transform. This is due to the connection between plane waves and Fourier harmonics in \eqref{2D-plane-wave}.

Accordingly, \eqref{source_spectrum} is the Fourier plane-wave transform of the source density. Notably, only the wavenumber points at $\kappa_z =  \pm \gamma$ contribute to the plane wave spectrum in \eqref{incident_spectrum}, which reveals the lower dimensional nature of the channel. This was also pointed out in \cite{FranceschettiLandau}, stating that the world has only an apparent 3D informational structure, which is subject to a 2D representation.
Notably, the 3D spectra $J_\pm(\kappa_x,\kappa_y)$ of the source density is parametrized by $(\kappa_x,\kappa_y)\in\Real^2$, which means the volumetric source can always be replicated exactly by a planar source of infinite extent. This in agreement with the fundamental Huygens principle in electromagnetic theory  and the physics Stokes' theorem \cite[Sec.~1.4]{ChewBook}.

Finally, it is worth mentioning that an alternative approach leading to \eqref{incident_field} may be followed. This involves computing the inverse spatial Fourier transform of \eqref{Helm_eq_wavenumber} with respect to $\kappa_z$. 
Due to the presence of real-valued poles in \eqref{wave_freq_response} for any lossless medium, integration must be performed in the complex $\kappa_z$ plane by using Cauchy's integral theorem and Jordan's lemma \cite{MarzettaIT}. However, some of the steps in \cite{MarzettaIT} are already included into the proof of the Weyl's identity (e.g., \cite[Sec.~2.2]{ChewBook}), which simplifies the analytical treatment.
 
 
\subsection{Fourier Plane-Wave Representation of Received Field}

%
%

While the transmitted field in \eqref{incident_field} is artificially created by the current density, a \emph{received field} $e_{\rm r}(\vect{r})$, upon interaction with the scatterers, is measured at any point $\vect{r}$. Clearly, $e_{\rm r}(\vect{r})$ does not require any external stimulus at the receiver to exist and is thus locally source-free. Physically, it must obey the homogeneous Helmholtz equation \cite[Sec.~1.2.2]{ChewBook}
\begin{equation}  \label{scalar_homo_Helm_eq}
\nabla^2 e_{\rm r}(\vect{r}) + \kappa^2 e_{\rm r}(\vect{r}) = 0
\end{equation}
which constitutes an eigenvalue equation of the Helmholtz operator $\left(\nabla^2 + \kappa^2  \right)$.
Natural eigen-solutions of \eqref{scalar_homo_Helm_eq} are the receive plane waves $e^{\imagunit (k_x r_x + k_y r_y \pm \gamma r_z)}$ \cite[Eq.~(1.2.23)]{ChewBook}.
The general solution to \eqref{scalar_homo_Helm_eq} is constructed by considering the entire eigenspace spanned by these eigenfunctions in the form of a 2D Fourier plane-wave representation \cite[Sec.~6.7]{StrattonBook}
\begin{align} \notag
e_{\rm r}(\vect{r})  & =  \iint_{-\infty}^\infty   \frac{dk_x}{2\pi}\frac{dk_y}{2\pi} \, e^{\imagunit (k_x r_x + k_y r_y)}  \Big(E_{\rm r}^+(k_x,k_y) e^{\imagunit \gamma r_z}  \\& \hspace{2cm} \label{received_field}
+ E_{\rm r}^-(k_x,k_y) e^{-\imagunit \gamma r_z}\Big)
 \end{align}
where each plane wave has arbitrary complex-valued amplitude $E_{\rm r}^\pm(k_x,k_y)$ for every received direction
\begin{align} \label{wavenumber_spherical_rx}
\hat\krx_{\pm} & = \hat{\vect{x}} \frac{k_x}{\kappa} + \hat{\vect{y}} \frac{k_y}{\kappa} \pm \hat{\vect{z}} \frac{\gamma(k_x,k_y)}{\kappa} \\& \label{wavenumber_spherical_rx_2} 
= \hat{\vect{x}} \sin\theta_{\rm r} \cos\phi_{\rm r} + \hat{\vect{y}} \sin\theta_{\rm r} \sin\phi_{\rm r}  + \hat{\vect{z}} \cos\theta_{\rm r}
\end{align}
given $(\theta_{\rm r},\phi_{\rm r}) \in [0,\pi] \times [0,2\pi)$ as elevation and azimuth angles in the receiver reference frame.
Similar to \eqref{incident_field}, the received field in \eqref{received_field} is created by an integral superposition of upgoing (i.e., $\theta_{\rm r}\in[0,\pi/2]$) and downgoing (i.e., $\theta_{\rm r}\in(\pi/2,\pi]$) plane waves (propagating and evanescent).
Differently to \eqref{incident_field}, where each plane-wave amplitude $E_{\rm t}^\pm(\kappa_x,\kappa_y)$ depends deterministically on $j(\vect{r})$ through \eqref{incident_spectrum}, the exact values of $E_{\rm r}^\pm(k_x,k_y)$ in \eqref{received_field} are generally not known as they may be related to $E_{\rm t}^\pm(\kappa_x,\kappa_y)$ through complicated interaction mechanisms. 

Summarizing, the sole action of a source density is to create a pair of transmitted plane-wave spectra $E_{\rm t}^\pm(\kappa_x,\kappa_y)$ (upgoing and downgoing) in each of the two half-spaces created by the source. An observer measures another pair of received plane-wave spectra $E_{\rm r}^\pm(k_x,k_y)$ (upgoing and downgoing). These four possible connections will be studied next in its most general form.

\subsection{Linear Scattering Operator} \label{sec:propagation_kernel}


\begin{figure*}[t!]
 \centering
  \includegraphics[width=.7\linewidth]{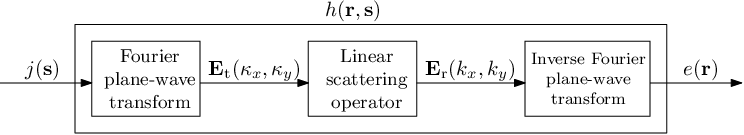}
\caption{Block diagram illustration of wave propagation in arbitrary NLoS channels.}
\label{fig:LSV_Fourier}
\end{figure*}  

Each plane wave from receive propagation direction $\hat\krx_{\pm}$ is the result of an interaction between \emph{all} transmit plane waves traveling towards $\hat\ktx_{\pm}$ and the scatterers. In other words, each receive plane-wave amplitude $E_{\rm r}^\pm(k_x,k_y)$ is induced by all transmitted plane-wave amplitudes $E_{\rm t}^\pm(\kappa_x,\kappa_y)$ for $(\kappa_x,\kappa_y)\in\Real^2$ through an integral functional operator modeling propagation in arbitrary environments \cite{Saxon,Gerjuoy,Kerns1976,NietoWolf}.
There are four possible physical interactions among the two plane-wave spectra created at source and other two spectra measured at receiver with associated functionals, which calls for a more suitable vector formulation. 
To this end, we stack upgoing and downgoing quantities associated with both fields together into a column vector.
At the source, we define for all $(\kappa_x,\kappa_y)$ a complex-valued \emph{transmit plane-wave spectrum}
\begin{equation} \label{source_vector}
\vect{E}_{\rm t}(\kappa_x,\kappa_y) = 
\begin{pmatrix}
E_{\rm t}^+(\kappa_x,\kappa_y) \\
E_{\rm t}^-(\kappa_x,\kappa_y)
\end{pmatrix}
\end{equation}
and an \emph{array response vector}
\begin{equation} \label{array_response}
\vect{a}(\kappa_x,\kappa_y,\vect{s}) = 
\begin{pmatrix}
e^{-\imagunit (\kappa_x s_x + \kappa_y s_y + \gamma(\kappa_x,\kappa_y) s_z)} \\
e^{-\imagunit (\kappa_x s_x + \kappa_y s_y - \gamma(\kappa_x,\kappa_y) s_z)}
\end{pmatrix}
\end{equation}
where the latter is known a-priori as depend uniquely on the source geometry.
Similarly, at receiver, \eqref{received_field} becomes
\begin{align} \label{received_field_vector}
e(\vect{r})  & =  \frac{1}{(2\pi)^2} \iint_{-\infty}^\infty  dk_x dk_y \, \vect{a}^{\Htran}(k_x,k_y,\vect{r})  \vect{E}_{\rm r}(k_x,k_y)
 \end{align}
with complex-valued \emph{receive plane-wave spectrum}
\begin{equation} \label{receive_vector}
\vect{E}_{\rm r}(k_x,k_y) = 
\begin{pmatrix}
E_{\rm r}^+(k_x,k_y) \\
E_{\rm r}^-(k_x,k_y)
\end{pmatrix}
\end{equation}
and associated array response vector $\vect{a}(k_x,k_y,\vect{r})$ obtained from \eqref{array_response}.
The entire informational structure of the interaction mechanism is contained in the wavenumber domain. Precisely, in a $2\times 2$ matrix functional mapping $\vect{E}_{\rm t}(\kappa_x,\kappa_y)$ in \eqref{source_vector} onto $\vect{E}_{\rm r}(k_x,k_y)$ in \eqref{receive_vector}.
Due to linearity of the Helmholtz equation in \eqref{Helm_eq} (and Maxwell's equation in general), we expect this map to be linear given by a complex integral operator $\vect{E}_{\rm r} = (\mathbb{K} \vect{E}_{\rm t})(k_x,k_y)$ defined as \cite{Saxon,Gerjuoy}
\begin{align} \label{receive_spectrum_vector} 
\vect{E}_{\rm r}(k_x,k_y) & = \iint_{-\infty}^\infty \!\! d\kappa_x d\kappa_y \, \vect{K}(k_x,k_y,\kappa_x,\kappa_y) \vect{E}_{\rm t}(\kappa_x,\kappa_y) 
\end{align}
where
\begin{align} \notag
&\vect{K}(k_x,k_y,\kappa_x,\kappa_y) = \\& \hspace{1cm} \label{propagation_kernel}
\begin{pmatrix}
K_{++}(k_x,k_y,\kappa_x,\kappa_y) & K_{+-}(k_x,k_y,\kappa_x,\kappa_y) \\
K_{-+}(k_x,k_y,\kappa_x,\kappa_y) & K_{--}(k_x,k_y,\kappa_x,\kappa_y) 
\end{pmatrix}
\end{align}
is the \emph{propagation kernel matrix} of the operator $\mathbb{K}$ mapping transmitted plane waves to every received plane wave. We keep the same sign convention used elsewhere in this paper for the entries of \eqref{propagation_kernel} being associated with upgoing and downgoing waves. Precisely, the first subscript refers to transmitted plane waves, while the second subscript refers to received plane waves; see Fig.~\ref{fig:propagation}. For example, the subscript $+-$ stands for upgoing transmitted plane waves and downgoing received plane waves. 
Also, since the propagation medium shall never amplify the transmitted field, every entries of \eqref{propagation_kernel} must be a square-integrable kernel such that
\begin{equation} \label{kernel_cauchy}
\iiiint_{-\infty}^\infty dk_x dk_y d\kappa_x d\kappa_y \,  |K_{\pm\pm}(k_x,k_y,\kappa_x,\kappa_y)|^2 \le 1
\end{equation}
which implies conservation of energy for any source of finite energy via Cauchy-Schwarz inequality. Equality in \eqref{kernel_cauchy} is achieved when the propagation medium is lossless so that no radiated energy is lost during transmission.  

Despite our formulation is applicable to every possible propagation scenario, a particular choice of the propagation kernel customizes the developed model to a prescribed environmental class (e.g., rural, urban, canyon). 
To this purpose, we next provide a few simplified examples on how to model $\vect{K}(k_x,k_y,\kappa_x,\kappa_y)$ in \eqref{propagation_kernel}.

\begin{example}[LoS propagation]
With no scatterers, all transmitted plane waves reach the receiver unaltered so that 
\begin{equation} \label{kernel_LoS}
\vect{K}(k_x,k_y,\kappa_x,\kappa_y) = (2\pi)^2 \delta(k_y-\kappa_y)\delta(k_x-\kappa_x) 
\begin{pmatrix}
1 & 0\\
0 & 0
\end{pmatrix}
\end{equation}
for all $r_z>R_0$.
The impulsive nature of this kernel is due to the one-to-one correspondence $\hat\krx_+ = \hat\ktx_+$ between every pair of source and receive directions. 
\end{example}

\begin{example}[Propagation via reflection]
Consider an infinite, $z-$oriented, smooth surface of arbitrary composition ($\epsilon,\mu$) that is located at $r_z=d_1$. The reflected channel created by the interaction with this surface is modeled as \cite{PizzoVTC22}
\begin{align} \notag
 \vect{K}(k_x,k_y,\kappa_x,\kappa_y) & = (2\pi)^2 \delta(k_y-\kappa_y)\delta(k_x-\kappa_x) \\& \label{kernel_reflection}  \hspace{1cm} R(k_x,k_y) e^{\imagunit 2 \gamma d_1}
\begin{pmatrix}
0 & 1\\
0 & 0
\end{pmatrix}
\end{align}
for all $d_1>R_0$. Here, $R(k_x,k_y)$ is the Fresnel reflection coefficient that specifies the fraction of transmitted plane-wave spectrum that is reflected off by the surface \cite[Sec.~2.1.2]{ChewBook}. The phase term in \eqref{kernel_reflection} accounts for the round-trip phase delay accumulated by each transmitted plane wave during its travel to the surface and backwards, along the $z$-axis.
\end{example}

\begin{example}[Multi-path propagation]
With scatterers located in the far field regions of source and receiver, propagation is specified by a finite number of preferred directions (or paths) generated by a cluster $i$. Each path $j$ is associated with a transmitted plane wave to direction $\hat \ktx_{+,j}$ and another receive plane wave from direction $\hat \krx_{+,j}$. Both are related to a complex propagation coefficient $K_{i,j}$. Altogether, for scatterers separating source and receiver,
\begin{align} \notag
& \vect{K}(k_x,k_y,\kappa_x,\kappa_y) =  (2\pi)^4 \sum_{i=1}^{N_{\rm c}} \sum_{j\in\mathcal{C}_i} K_{i,j}  \, 
\delta(k_y-k_{y,j}) \\& \label{ray_tracing} \hspace{1cm}\delta(k_x-k_{x,j}) 
\delta(\kappa_y-\kappa_{y,j})\delta(\kappa_x-\kappa_{x,j})  
\begin{pmatrix}
1 & 0\\
0 & 0
\end{pmatrix}
\end{align}
where $N_{\rm c}$ is the number of clusters and $\mathcal{C}_i$ the number of paths within each cluster. 
\end{example}

In general, the interaction mechanism between the plane-wave spectra at source and receiver will be given by a composition of different physical phenomena. Nevertheless, due to linearity of the scattering operator, each of these physical interactions can be modeled independently by specifying a suitable kernel matrix $\vect{K}(k_x,k_y,\kappa_x,\kappa_y)$ and finally  added together in \eqref{receive_spectrum_vector} to obtain the total contribution. 

\subsection{Impulse Response of NLoS Electromagnetic Channels}

The input-output relationship between $e(\vect{r})$ and $j(\vect{s})$ is the spatial convolution:
\begin{align}  \label{channel_NLoS}
e(\vect{r}) = \int_{\Real^3} d\vect{s} \, j(\vect{s}) h(\vect{r},\vect{s})
\end{align}
where $h(\vect{r},\vect{s})$ is the space-variant channel impulse response at point $\vect{r}$ due to a unit impulse (point source) applied at point $\vect{s}$. Compared to the LoS scenario, the space variance of the channel is caused by the propagation environment now being sensitive to a space-shift of source and/or receiver (i.e., the relative distances and angles among source, receiver, and scatterers change). The closed-form expression of $h(\vect{r},\vect{s})$ is provided next. 

\begin{theorem} \label{th:4D_plane_wave_representation_complete}
The channel response $h(\vect{r},\vect{s})$ modeling an arbitrary NLoS propagation environment is exactly given by the 4D Fourier plane-wave representation 
\begin{align} \notag
h(\vect{r},\vect{s}) &  =  \frac{1}{(2\pi)^2}
 \iiiint_{-\infty}^\infty  dk_x dk_y d\kappa_x d\kappa_y \, \vect{a}^{\Htran}(k_x,k_y,\vect{r}) \\& \hspace{2cm} \label{channel_response_complete} 
 \vect{H}(k_x,k_y,\kappa_x,\kappa_y) 
   \vect{a}(\kappa_x,\kappa_y,\vect{s})
\end{align}
where $\vect{a}(\cdot,\cdot)$ is the array response vector in \eqref{array_response} and we introduced the angular response matrix
\begin{equation} \label{angular_response_complete}
\vect{H}(k_x,k_y,\kappa_x,\kappa_y) =  \frac{\kappa \eta}{2} \frac{\vect{K}(k_x,k_y,\kappa_x,\kappa_y)}{\sqrt{\gamma(k_x,k_y)} \sqrt{\gamma(\kappa_x,\kappa_y)}}
\end{equation}
parametrized by $\vect{K}(k_x,k_y,\kappa_x,\kappa_y)$ in \eqref{propagation_kernel}.
\end{theorem}
\begin{proof}
The proof is given in Appendix~\ref{app:channel_response}. 
 \end{proof}

The above result generalizes the channel impulse response in \eqref{impulse_response_LoS_plane_Wave}, derived for a LoS scenario, to arbitrary propagation environments. The LoS channel is obtainable from \eqref{channel_response_complete} after substituting the corresponding propagation kernel in \eqref{kernel_LoS}.

We can breakdown \eqref{channel_response_complete} as generated by three contributions. 
The first term is the array response vector $\vect{a}(\kappa_x,\kappa_y,\vect{s})$ in \eqref{array_response}  that maps an impulsive excitation current at $\vect{s}$ to every outgoing propagation direction $\hat\ktx_{\pm}$. Similarly, the second term is the array response vector $\vect{a}^{\Htran}(k_x,k_y,\vect{r})$ that maps every incoming propagation direction $\hat\krx_{\pm}$ to the induced current at $\vect{r}$. The third term $\vect{H}(k_x,k_y,\kappa_x,\kappa_y)$ of the channel $h(\vect{r},\vect{s})$ is given in \eqref{angular_response_complete} and represents the \emph{angular response matrix} that maps every incident direction $\hat\ktx_{\pm}$ into every other receive direction $\hat\krx_{\pm}$, for all four combinations of upgoing and downgoing directions.
Clearly, the sole action of an array response is to change domain of representation, i.e., from spatial to angular and vice-versa. Both contain information about the source and receiver geometry. Everything else is embedded into the angular response that fully describes the underlying interaction mechanism through \eqref{propagation_kernel}. 

Remarkably, the separate structure of $h(\vect{r},\vect{s})$ in \eqref{channel_response_complete} is enforced by physics without any additional assumption. Suitably sampled in the spatial domain, it leads to a decoupled structure of the MIMO channel matrix where impact of array configuration and scatterers are separated. This property can be leveraged to build a transceiver architecture where antenna placement and signal processing algorithms are designed  independently \cite{PizzoTWC21}. Also, the number of radio frequency chains required for data processing is fundamentally lower than the number of antennas used at the front end, as it depends on the solid angle subtended by the scatterers and array size jointly. Altogether brings a significant complexity reduction in channel estimation, optimal signaling, and coding \cite{PizzoTWC21}.

\subsection{Wavenumber Response}

For any fixed pair $(r_z,s_z)$, each entry of \eqref{array_response} correspond to a phase-shifted version of a 2D spatial-frequency Fourier harmonic, where the phase shift is applied along the $z$-axis,
\begin{equation} \label{2D-plane-wave_vector}
\vect{a}(\kappa_x,\kappa_y,\vect{s})  = e^{-\imagunit \left(\kappa_x s_x + \kappa_y s_y \right)}  \boldsymbol{\phi}(\kappa_x,\kappa_y,s_z)
\end{equation}
with $\boldsymbol{\phi}(\kappa_x,\kappa_y,s_z) = [e^{-\imagunit  \gamma(\kappa_x,\kappa_y) s_z}, e^{\imagunit  \gamma(\kappa_x,\kappa_y) s_z}]^{\Ttran} $.
Building on this analogy, $h(\vect{r},\vect{s})$ in \eqref{channel_response_complete} may be rewritten as a 4D inverse spatial Fourier transform\footnote{The change of sign in \eqref{impulse_response_Fourier} at the receiver, with respect to the standard Fourier transform, is due to a change of the reference system from source to receiver. See also Appendix~\ref{app:reciprocity}.}
\begin{align} \notag
h(\vect{r},\vect{s}) =  & \frac{1}{(2\pi)^2}
 \iiiint_{-\infty}^\infty  dk_x dk_y d\kappa_x d\kappa_y \,e^{\imagunit \left(k_x r_x + k_y r_y \right)} \\& \hspace{1.5cm} \label{impulse_response_Fourier} H(k_x,k_y,\kappa_x,\kappa_y)  e^{-\imagunit \left(\kappa_x s_x + \kappa_y s_y \right)}
\end{align}
in terms of its wavenumber response
\begin{align} \notag
&H(k_x,k_y,\kappa_x,\kappa_y) =\\& \hspace{1cm} \boldsymbol{\phi}^{\Htran}(k_x,k_y,r_z)  \vect{H}(k_x,k_y,\kappa_x,\kappa_y)  \label{H_spectrum}  \boldsymbol{\phi}(\kappa_x,\kappa_y,s_z)
\end{align}
where the spectrum dependence on the parameters $(r_z,s_z)$ is omitted.
The obtained Fourier representation in \eqref{impulse_response_Fourier} should be compared to the plane-wave representation in \eqref{channel_response_complete}. While the former is the map between the {space domain}  and {angular domain} in terms of plane waves, the latter represents the map between the {space domain}  and {wavenumber (spatial-frequency) domain} in terms of Fourier harmonics.
The entire information of the channel is embedded into the wavenumber response $H(k_x,k_y,\kappa_x,\kappa_y)$ of the channel in \eqref{H_spectrum}, which specifies the response at spatial frequency $(k_x,k_y)$ due to an oscillating input at spatial frequency $(\kappa_x,\kappa_y)$. The upgoing-upgoing term in \eqref{wavenumber_response_LoS} is one of the four spectral contributions obtained by using \eqref{angular_response_complete} into \eqref{H_spectrum}.

Similarly to the LoS scenario, the channel is subjected to a filtering operation due to migration filters that determines the maximum available bandwidth \cite{PizzoSPAWC20,PizzoTSP21}. On the contrary, in a NLoS scenario, the available bandwidth depends on the richness of the scattering through the wavenumber support of the propagation kernel, say $\mathcal{K} \subseteq \mathcal{D}$, 
\begin{equation} 
m(\mathcal{K}) \le m(\mathcal{D})
\end{equation}
where the equality is achieved under isotropic propagation for which we have that $\mathcal{K} = \mathcal{D}$ (see also Section~\ref{sec:isotropic}). The above consideration can be used to determine the dimensionality of an electromagnetic channel under arbitrary NLoS conditions, namely the degrees of freedom (DoF). In particular, the number of DoF per unit area is derived from Landau's eigenvalue formula as $m(\mathcal{K})/(2\pi)^2$ \cite{PizzoTSP21}. This result is as tight as the array size is large compared to the wavelength and has similar implications to the Shannon's DoF formula for time-domain channels.

 \subsection{Channel Reciprocity for Downlink Communications} \label{sec:reciprocity}

The behavior of a scalar electromagnetic channel is governed by the Helmholtz equation in \eqref{Helm_eq}. Rooted in the symmetry of the Helmholtz operator, the \emph{reciprocity theorem} states that the channel remains unchanged if one interchanges the points where the source is placed and the field is measured \cite[Sec.~3.8]{BalanisBook}. For a scalar LoS channel, it yields reciprocity of the Green's function in \eqref{Green}, i.e., $G(\vect{r},\vect{s}) = G(\vect{s},\vect{r})$. Notice that this is a stronger condition than the space-invariance of $h(\vect{r}-\vect{s})$ and it is due to the rotational symmetry of \eqref{Green}.
Generalization to a vector electromagnetic channel (i.e., including polarization) implies an additional symmetry of the dyadic Green's function matrix \cite[Sec.~1.3.4]{ChewBook}.

Building upon this, we apply reciprocity theorem to arbitrary NLoS channels and look at its implication on the angular domain. The main result is summarized next.

\begin{lemma} \label{th:reciprocity}
Let $h(\vect{r},\vect{s})$ be the channel response of an arbitrary NLoS channel system given by Theorem~\ref{th:4D_plane_wave_representation_complete}. The response $h(\vect{s},\vect{r})$ of the reciprocal system obtained interchanging source and receiver has the same form, but with angular response matrix $\vect{H}^{\Ttran}(-\kappa_x,-\kappa_y,-k_x,-k_y)$.
\end{lemma}
\begin{proof}
The proof is given in Appendix~\ref{app:reciprocity}. 
 \end{proof}

Intuitively, interchanging source and receiver while maintaining the same reference system, implies reversing the direction of propagation. This operation has a three-fold effect on the angular response matrix: the interchange between source and receive propagation directions with change of variables from $(k_x,k_y)$ to $(\kappa_x,\kappa_y)$ and vice versa, a reflection of all propagation directions about the $z$-axis with additional negative sign from $(k_x,k_y)$ to $(-\kappa_x,-\kappa_y)$, and a transpose operation due to upgoing (downgoing) plane waves becoming downgoing (upgoing) plane waves. As an example of channel reciprocity, we next provide the propagation kernel leading to the reciprocal channel for the  examples provided in Section~\ref{sec:propagation_kernel}. For the LoS case, we obtain
\begin{equation} \label{kernel_LoS_reciprocal}
\vect{K}(k_x,k_y,\kappa_x,\kappa_y) = (2\pi)^2 \delta(k_y-\kappa_y)\delta(k_x-\kappa_x) 
\begin{pmatrix}
0 & 0\\
0 & 1
\end{pmatrix}
\end{equation}
for all $r_z<-R_0$. With respect to \eqref{kernel_LoS}, we notice a downgoing-downgoing interaction only in the scattering matrix.  
Instead, for the reflected channel from a $z-$oriented half-space,
\begin{align} \notag
\vect{K}(k_x,k_y,\kappa_x,\kappa_y) &= (2\pi)^2 \delta(k_y-\kappa_y)\delta(k_x-\kappa_x) \\& \hspace{1cm}\label{kernel_reflection} R(-k_x,-k_y) e^{\imagunit 2 \gamma d_1}
\begin{pmatrix}
0 & 0\\
1 & 0
\end{pmatrix}
\end{align}
for all $d_1<-R_0$. 


\section{Stochastic characterization of an electromagnetic channel} \label{sec:statistical_model}

\begin{figure*}
\begin{align}  \notag
& h(\vect{r},\vect{s})  =  \frac{1}{(2\pi)^2}
 \iiiint_{\mathcal{D} \times \mathcal{D}}  dk_x dk_y d\kappa_x d\kappa_y \, \frac{e^{\imagunit \left(k_x r_x + k_y r_y \right)}}{\sqrt{\gamma(k_x,k_y)}} \frac{e^{-\imagunit \left(\kappa_x s_x + \kappa_y s_y \right)}}{\sqrt{\gamma(\kappa_x,\kappa_y)}}  \\ \notag & \hspace{6cm}
 \Big(A_{++}(k_x,k_y,\kappa_x,\kappa_y) W_{++}(k_x,k_y,\kappa_x,\kappa_y) e^{-\imagunit  \gamma(\kappa_x,\kappa_y) s_z} e^{\imagunit  \gamma(k_x,k_y) r_z} +  \\ \notag & \hspace{6cm}
 A_{+-}(k_x,k_y,\kappa_x,\kappa_y) W_{+-}(k_x,k_y,\kappa_x,\kappa_y) e^{-\imagunit  \gamma(\kappa_x,\kappa_y) s_z} e^{-\imagunit  \gamma(k_x,k_y) r_z} + \\ \notag & \hspace{6cm}
 A_{-+}(k_x,k_y,\kappa_x,\kappa_y) W_{-+}(k_x,k_y,\kappa_x,\kappa_y) e^{\imagunit  \gamma(\kappa_x,\kappa_y) s_z} e^{\imagunit  \gamma(k_x,k_y) r_z} + \\ & \hspace{6cm} \tag{55} \label{Fourier_spectral_random}
 A_{--}(k_x,k_y,\kappa_x,\kappa_y) W_{--}(k_x,k_y,\kappa_x,\kappa_y) e^{\imagunit  \gamma(\kappa_x,\kappa_y) s_z} e^{-\imagunit  \gamma(k_x,k_y) r_z} \Big)
\end{align}
\hrule
 \end{figure*}
 
Stochastic channel models have been used by communication theorists since their introduction due to their wide range of applicability \cite{Bello}. In a stochastic formulation, the channel response is modeled as a \emph{spatial electromagnetic random field} with each realization being representative of wave propagation into a hypothetically different environment. This is generally given by the sum of a deterministic component plus another zero-mean random component, which yields Rician fading. We next focus on the random component.
Small variations of the propagation environment are typically accounted by modeling $h(\vect{r},\vect{s})$ as a zero-mean circularly symmetric complex Gaussian random field, which yields a Rayleigh small-scale fading. Here, the Gaussian assumption arises as a {diffusion approximation} of the scattering mechanism \cite{Aris}. 
Moreover, we assume the second-order statistics of the channel are space invariant, leading to a substantial model simplification.
Hence, we will develop our stochastic model under two main assumptions: \emph{complex Gaussian distribution} and \emph{spatial stationarity}; $h(\vect{r},\vect{s})$ being a {spatially stationary complex Gaussian electromagnetic random field} \cite{MarzettaISIT}. 
Larger variations of the propagation environment should be modeled by a non-stationary large-scale fading field \cite[Sec.~7]{MolischBook}.

%

\subsection{Spatially Stationary Gaussian Random Channels}

Spatial stationarity is a desirable property for time-domain Gaussian random processes as they are fully characterized by a second-order description of their statistics.\footnote{We refer to a stationary process since strict stationarity and stationarity in the wide-sense coincide for any random process with joint Gaussian distribution.} For example, it is at the basis of the wide-sense stationary uncorrelated scattering (WSSUS) model for linear and time-variant channels \cite{Bello}. 
Let $h(t,\tau)$ be the time-variant channel impulse response at time $t$ due to a unit impulse applied at $t-\tau$ (with a delay $\tau$). A Gaussian random channel featuring a WSSUS model is characterized by a stationarity with respect to the variable $t$ and uncorrelated values at different delays $\tau$.

For the channel response $h(\vect{r},\vect{s})$, we assume spatial stationarity with respect to both spatial variables, so that a joint autocorrelation function (ACF) can be defined as
\begin{equation} \label{autocorrelation}
c(\vect{r}, \vect{s}) = \Ex\{h(\vect{r},\vect{s}) \, h^*(\vect{r}+\vect{r}^\prime,\vect{s}+\vect{s}^\prime)\}.
\end{equation}
We will now show that both assumptions on $h(\vect{r},\vect{s})$ lead to an \emph{independent scattering} model for the propagation kernel $\vect{K}(k_x,k_y,\kappa_x,\kappa_y)$ whose structure is given next.
        
\begin{theorem} \label{th:stationary}
The propagation kernel matrix $\vect{K}(k_x,k_y,\kappa_x,\kappa_y)$ leading to a spatially stationary circularly symmetric  complex Gaussian random $h(\vect{r},\vect{s})$ in \eqref{channel_response_complete} must be of the form 
\begin{equation} \label{angular_response_stationary}
\vect{K}(k_x,k_y,\kappa_x,\kappa_y) =  \vect{A}(k_x,k_y,\kappa_x,\kappa_y) \odot \vect{W}(k_x,k_y,\kappa_x,\kappa_y)
\end{equation}
where $\vect{A}(k_x,k_y,\kappa_x,\kappa_y) \in\Real_+^{2\times 2}$ whose entries are arbitrary non-negative functions defined within $(k_x,k_y,\kappa_x,\kappa_y)\in\mathcal{D}\times\mathcal{D}$ with $\mathcal{D}$ given by~\eqref{disk_T} and $\vect{W}(k_x,k_y,\kappa_x,\kappa_y) \in\Complex^{2\times 2}$ is a random matrix with white-noise complex random field entries of unit-variance,
\begin{align} \notag
& \Ex\{[\vect{W}(k_x,k_y,\kappa_x,\kappa_y)]_{i,\ell} [\vect{W}^*(k_x^\prime,k_y^\prime,\kappa_x^\prime,\kappa_y^\prime)]_{i^\prime,\ell^\prime}\} = \delta_{i i^\prime} \delta_{\ell \ell^\prime} \\ \label{white_noise_matrix} & \hspace{.8cm}  \delta(k_x - k_x^\prime) \delta(k_y - k_y^\prime) \delta(\kappa_x - \kappa_x^\prime) \delta(\kappa_y - \kappa_y^\prime).
\end{align}
\end{theorem}
\begin{proof}
The proof is given in Appendix~\ref{app:stationarity} and is articulated in two parts involving Gaussianity and stationarity.
\end{proof}

The propagating kernel $\vect{K}(k_x,k_y,\kappa_x,\kappa_y)$ in \eqref{angular_response_stationary} is obtained as an element-wise multiplication of two terms. The first term $\vect{A}(k_x,k_y,\kappa_x,\kappa_y)$ is a non-negative real-valued matrix, which models directionality of the channel field. Precisely, it physically accounts for the angular power transfer between every upgoing (downgoing) transmitted direction and every other upgoing (downgoing) received direction, averaged over all possible realizations of a certain environmental class; see Fig.~\ref{fig:propagation}. This is the only functional parameter that must be specified in our model. We will return to this later on in Section~\ref{sec:phy_model_spectral}.
The second term $\vect{W}(k_x,k_y,\kappa_x,\kappa_y)$ is a random matrix whose objective is to allow for small variations among different realizations. 
Both model microscopic effects on the channel small-scale fading caused by small changes in the propagation environment.
Combined together, Theorem~\ref{th:4D_plane_wave_representation_complete} and Theorem~\ref{th:stationary} yield the following closed-form expression of $h(\vect{r},\vect{s})$.

 \begin{lemma} \label{th:channel_response_random}
 The random channel impulse response $h(\vect{r},\vect{s})$ modeling an arbitrary propagation environment is exactly given by the  Fourier plane-wave representation in \eqref{Fourier_spectral_random}. Here, $A_{\pm\pm}(k_x,k_y,\kappa_x,\kappa_y)$
 are four non-negative functions defined within a support $\mathcal{D}$ given by~\eqref{disk_T} and $W_{\pm\pm}(k_x,k_y,\kappa_x,\kappa_y)$ are four i.i.d. white-noise complex random fields of unit variance.
 \end{lemma}
   
A physical interpretation of the result reported in Lemma~\ref{th:channel_response_random} is as follows.
The channel impulse response modeling a spatially stationary random medium is obtained as an integral superposition of (upgoing and downgoing) \emph{propagating} plane waves having statistically independent amplitudes from one direction to another and jointly having circularly symmetric complex Gaussian distribution. 
Alternatively, due to the interchangeability between plane waves and Fourier harmonics, \eqref{Fourier_spectral_random} can also be regarded as the Fourier spectral representation of a stationary random field of electromagnetic nature, returning a  Fourier description similar to \eqref{impulse_response_Fourier}. This will be shown in Section~\ref{sec:spectral_representation}.

As stated in Appendix~\ref{app:stationarity}, spatial stationarity requires the exclusion of evanescent waves from our analysis. These are associated to the high wavenumber modes outside of $\mathcal{D}$ that are generated by the source and possibly by induced currents on the surface of scatterers. 
Since these modes decay exponentially fast as $z/\lambda$, their contribution may be neglected at a few wavelengths from the radiators. Notice that the same property was inspected in Corollary~\ref{th:bandlimited} for LoS channels and is extended here to arbitrary NLOS channels.
The downside is that some of the available channel information is lost in this low-pass filtering operation \cite{PizzoTSP21}. Consequently, wireless transfer of information is always a lossy operation with the majority of communication modes wasted in the reactive propagation region of source and scatterers \cite{PizzoSPAWC20,PizzoTSP21}.

\subsection{Second-Order Characterization of Stationary Channels}

\begin{figure*}
\begin{align} \tag{58}\label{autocorrelation3}
c(\vect{r}, \vect{s}) & =\frac{1}{(2\pi)^3} \idotsint_{-\infty}^\infty S(k_x,k_y,k_z,\kappa_x,\kappa_y,\kappa_z) e^{\imagunit (k_x r_x + k_y r_y + k_z r_z)}  
e^{-\imagunit (\kappa_x s_x + \kappa_y s_y + \kappa_z s_z)} dk_x dk_y dk_z d\kappa_x d\kappa_y d\kappa_z
\end{align}
\hrule
\end{figure*}

 \begin{figure} [t!]
        \centering
	\begin{overpic}[width=.99\columnwidth,tics=10]{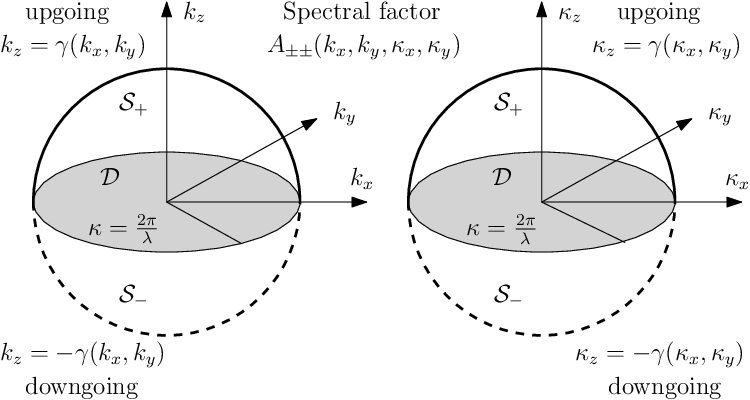}
\end{overpic} \vspace{-0.0cm}
                \caption{Illustration of the PSD $S(k_x,k_y,k_z,\kappa_x,\kappa_y,\kappa_z)$ of a spatially stationary channel impulse response $h(\vect{r},\vect{s})$, fully parametrized by the four non-negative functions $A_{\pm \pm}(k_x,k_y,\kappa_x,\kappa_y)$ defined within $\mathcal{D} \times \mathcal{D}$. 
                } \vspace{0.0cm}
                \label{fig:spectral_distribution} 
        \end{figure} 
        
For a stationary Gaussian random field $h(\vect{r},\vect{s})$, the spatial ACF $c(\vect{r}, \vect{s})$ in \eqref{autocorrelation} provides a complete statistical description of the field. Its general expression for arbitrary propagation environments is provided next. Similarly to $h(\vect{r};\vect{s})$ in~\eqref{channel_response_complete}, $c(\vect{r}, \vect{s})$ is also described by a Fourier plane-wave representation. This comes with no surprise as the ACF of the channel also satisfies the Helmholtz equation \cite{MarzettaISIT,PizzoJSAC20}. 

\begin{lemma} \label{th:correlation_function}
The spatial ACF $c(\vect{r}, \vect{s})$ of the channel impulse response $h(\vect{r},\vect{s})$ modeling an arbitrary propagation environment is exactly given by
\setcounter{equation}{55} 
\begin{align} \notag
& c(\vect{r}, \vect{s})  =  \frac{1}{(2\pi)^2}
 \iiiint_{-\infty}^\infty dk_x dk_y d\kappa_x d\kappa_y \, \vect{a}^{\Htran}(k_x,k_y,\vect{r}) \\&  \hspace{3cm}\label{autocorrelation_Fourier_plane_wave}  \vect{S}(k_x,k_y,\kappa_x,\kappa_y)  
    \vect{a}(\kappa_x,\kappa_y,\vect{s})
\end{align}
where $\vect{a}(\cdot,\cdot)$ is the array response vector in \eqref{array_response} and we introduced the power density matrix\footnote{The channel response under spatial stationarity has constant power. Hence, we remove the multiplicative term $(\kappa \eta/2)^2$ that would have appeared in \eqref{spectral_matrix} and embedded it into the spectral factor.}
\begin{equation} \label{spectral_matrix}
\vect{S}(k_x,k_y,\kappa_x,\kappa_y) = \frac{(\vect{A}\odot \vect{A})(k_x,k_y,\kappa_x,\kappa_y)}{\gamma(k_x,k_y) \gamma(\kappa_x,\kappa_y)}
\end{equation}
which is parametrized by $\vect{A}(k_x,k_y,\kappa_x,\kappa_y)$ in \eqref{angular_response_stationary}.
\end{lemma}
\begin{proof}
Substitute the Fourier plane-wave representation in Theorem~\ref{th:4D_plane_wave_representation_complete} into \eqref{autocorrelation}. Then, use the second order characterization of the propagation kernel in Theorem~\ref{th:stationary}.
\end{proof}

The standard six-dimensional power spectral density (PSD)  $S(k_x,k_y,k_z,\kappa_x,\kappa_y,\kappa_z)$, function of the three spatial frequencies at source plus other three spatial frequencies at receiver, is related to $c(\vect{r}, \vect{s})$ through the Fourier relationship in \eqref{autocorrelation3} due to Wiener-Kintchine theorem. In \eqref{autocorrelation3}, we have changed the sign of Fourier harmonics at the source -- with respect to the ordinary inverse Fourier transform -- to comply with the notation used in this paper. The general form that the PSD of a stationary channel must have under arbitrary propagation conditions was determined in \cite{MarzettaISIT,PizzoJSAC20} for the source-free case at receiver.  We extend that result to an end-to-end propagation scenario including a source.
\setcounter{equation}{54} 
\begin{lemma} \label{th:psd}
The PSD of any spatially stationary channel impulse response $h(\vect{r},\vect{s})$ is impulsive of the form 
\setcounter{equation}{58} 
\begin{align} \notag
 & S(k_x,k_y,k_z,\kappa_x,\kappa_y,\kappa_z)  =  A^2(k_x,k_y,k_z,\kappa_x,\kappa_y,\kappa_z) \\& \label{psd_final} \hspace{1.5cm}
\delta(k_x^2 + k_y^2 + k_z^2- \kappa^2) \delta(\kappa_x^2 + \kappa_y^2 + \kappa_z^2- \kappa^2)
\end{align}
for some non-negative function $A(k_x,k_y,k_z,\kappa_x,\kappa_y,\kappa_z)$. 
\end{lemma}
\begin{proof}
The proof is given in Appendix~\ref{app:psd}. 
\end{proof}


The PSD in~\eqref{autocorrelation3} is impulsive defined on a double sphere $\mathcal{S}\subset \Real^3$ of radius $\kappa=2\pi/\lambda$. Clearly, this is due to the Fourier (plane-wave) description of the spatial ACF that yields a pair of Dirac delta functions in the wavenumber dual domain, at source and receiver. The term $A(k_x,k_y,k_z,\kappa_x,\kappa_y,\kappa_z)$ in \eqref{psd_final} describes how the channel power is distributed over the spectral support, called the \emph{spectral factor}. Clearly, these six-dimensional quantities shall never be used since they are only meaningful when integrated. A 4D second-order representation should be used instead. This is provided in Lemma~\ref{th:correlation_function} wherein each entry $A_{\pm \pm}(k_x,k_y,\kappa_x,\kappa_y)$ of $\vect{A}(k_x,k_y,\kappa_x,\kappa_y)$ is obtained by sampling the spectral factor in the wavenumber domain at $k_z = \pm \gamma(k_x,k_y)$ and $\kappa_z = \pm \gamma(\kappa_x,\kappa_y)$, as illustrated in Fig.~\ref{fig:spectral_distribution} (see also Appendix~\ref{app:psd}). 
As previously observed in \cite{PizzoJSAC20}, due to the Dirac delta functions, the above sampling operation corresponds to a wavenumber integration of the PSD over $k_z$ and $\kappa_z$. Precisely, we divide the two spectral spheres $\mathcal{S}$ into four hemispheres -- two hemispheres (upper and lower) at source and other two hemispheres (upper and lower) at receiver. Each 3D hemisphere is then parametrized onto the corresponding 2D disk $\mathcal{D}$ in \eqref{disk_T}; see Fig.~\ref{fig:spectral_distribution}. To this regard, the $\gamma(\cdot,\cdot)$ functions in \eqref{spectral_matrix} are the Jacobians of these parametrizations.

The average channel power $P = \Ex\{|h(\vect{r},\vect{s})|^2\} $ is obtained by integrating the PSD in \eqref{psd_final} over its entire support. Alternatively, $P$ may be derived from \eqref{autocorrelation_Fourier_plane_wave} by sampling the array response vectors in \eqref{array_response} at the origin, 
 \begin{align}\label{power}
P & = \frac{1}{(2\pi)^4} \iiiint_{-\infty}^\infty  S(k_x,k_y,\kappa_x,\kappa_y)  \, dk_x dk_y d\kappa_x d\kappa_y
\end{align}
where $S(k_x,k_y,\kappa_x,\kappa_y)$ is obtained by summing all four entries of $\vect{S}(k_x,k_y,\kappa_x,\kappa_y)$ in \eqref{spectral_matrix}. The above expression motivates the name attributed to $\vect{S}(k_x,k_y,\kappa_x,\kappa_y)$ in Lemma~\ref{th:correlation_function}, being the power density matrix of the channel. This is because, when integrated over the four horizontal wavenumber components, the sum of its entries yields the channel power.

\section{Fourier spectral representation} \label{sec:spectral_representation}

Signals can be represented as an integral superposition of complex oscillations via the classical inverse Fourier transform. Similarly, stationary random processes may be represented as an integral superposition of uncorrelated complex oscillations, also known as the \emph{Fourier spectral representation} \cite[Ch.~4]{VanTreesBook}.
If, in addition, we require the process to have Gaussian distribution, these oscillations are statistically independent and the random process is fully described by its PSD function.
We show how this generalizes to spatially stationary Gaussian electromagnetic channels.

\subsection{Stationary Random Processes}  \label{sec:stationary_process}
 
The Fourier spectral representation of a stationary Gaussian random process $h(t)$ for $t\in\Real$ reads as \cite{VanTreesBook}
\begin{equation} \label{Fourier_spectral_rigorous}
h(t) = \int_{-\infty}^{\infty} e^{\imagunit \omega t} \,{dZ(\omega)}
\end{equation}
where the above equality must be understood to hold as a limit in mean-squared-error sense. 
Here, $Z(\omega) $ is the complex-valued integrated Fourier transform of $h(t)$ such that
\begin{equation} \label{integrated_spectrum}
\Ex\{dZ(\omega) dZ^*(\lambda)\} = 
\begin{cases}
dP(\omega)/2\pi, &  \lambda = \omega \\
0, & \text{otherwise}
\end{cases}
\end{equation}
with $dP(\omega)$ being the real-valued differential power increment of $h(t)$.
For any random process that contains no periodic terms, we have that  $dP(\omega) = S(\omega) d\omega$ where $S(\omega)$ is the real-valued and absolutely continuous PSD of $h(t)$. Hence, \eqref{Fourier_spectral_rigorous} can be rewritten in its Riemann form as  \cite{GrangerBook}
\begin{equation} \label{Fourier_spectral_continous}
h(t) = \frac{1}{\sqrt{2\pi}}\int_{-\infty}^\infty \sqrt{S(\omega)} W(\omega) e^{\imagunit \omega t} \,{d\omega}
\end{equation}
where $W(\omega)$ is a white-noise complex random process with unit variance. 
The ACF is obtained from the Wiener-Khintchine theorem as
\begin{equation} \label{acf_Fourier_spectral_continous}
c(t) = \frac{1}{2\pi}\int_{-\infty}^\infty S(\omega) e^{\imagunit \omega t} \,{d\omega}.
\end{equation}

\subsection{Spatially-Stationary Random Electromagnetic Channels}

 \begin{figure*}[th!] 
 \begin{subfigure}{0.33\textwidth}
 \hspace{-.2cm}
  \includegraphics[width=1.08\linewidth]{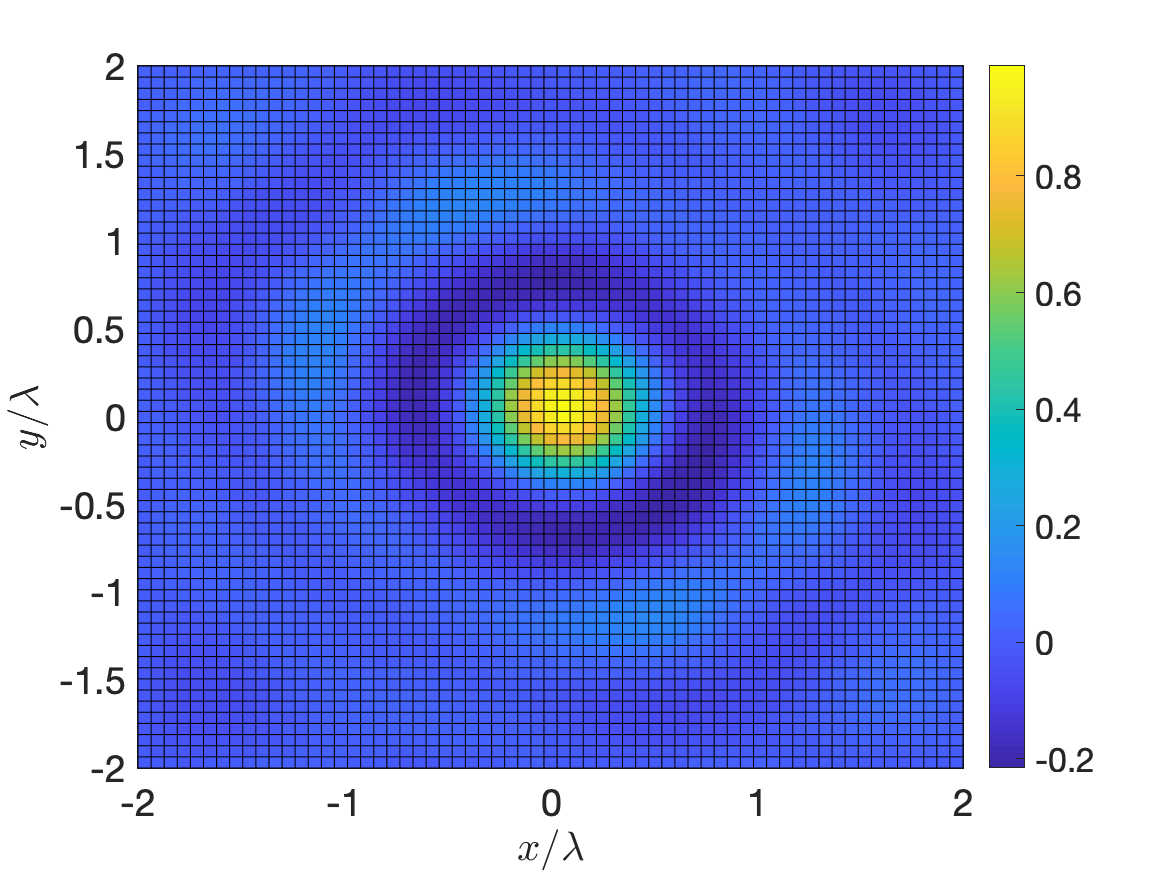}
  \caption{$L/\lambda=4$}\label{fig:eigenvalues_Iso}
\end{subfigure} 
\begin{subfigure}{0.33\textwidth}
\hspace{-.2cm}
  \includegraphics[width=1.08\linewidth]{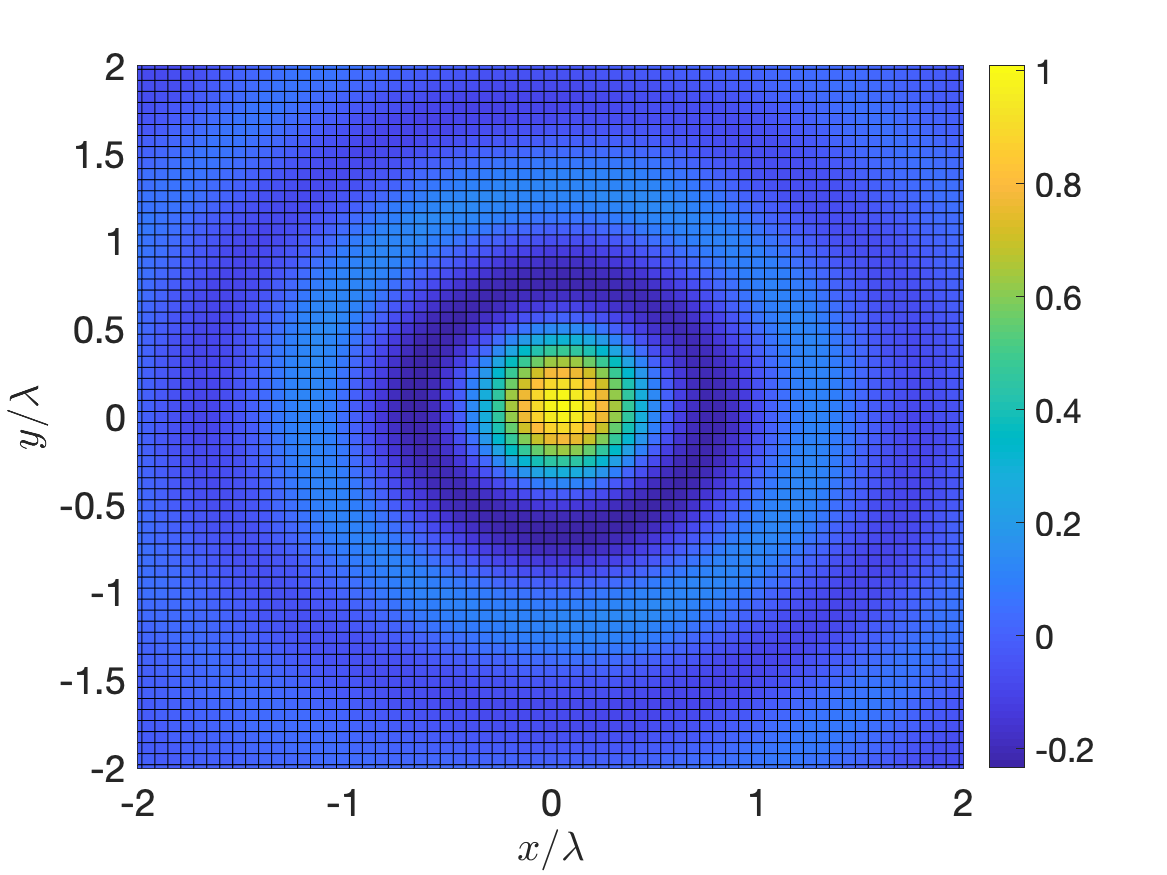}
  \caption{$L/\lambda=10$}\label{fig:eigenvalues_NIsoCPL}
\end{subfigure}
\begin{subfigure}{0.33\textwidth}%
\hspace{-.2cm}
  \includegraphics[width=1.08\linewidth]{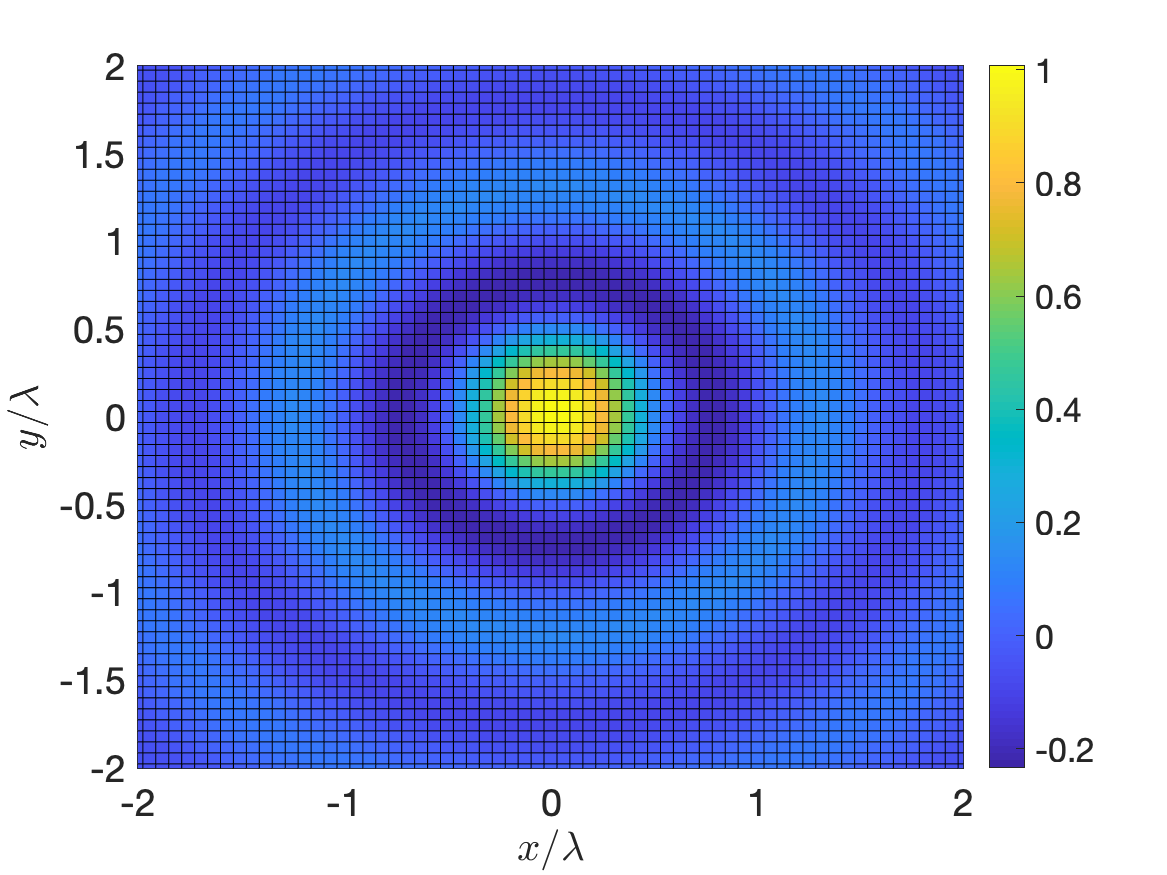}
  \caption{Closed-form $(L/\lambda\to\infty)$}\label{fig:eigenvalues_NIsoCluster}
\end{subfigure}
\caption{ACF at receiver under isotropic propagation. The Fourier plane-wave series expansion with different $L/\lambda$ values is compared to the closed-form Clarke's isotropic function.}
\label{fig:acf}
\end{figure*} 

The Fourier spectral representation of a stationary Gaussian random channel $h(\vect{r},\vect{s})$ for $\vect{r},\vect{s} \in \Real^3$ would be obtained similarly to \eqref{Fourier_spectral_continous} as a function of its six-dimensional PSD. However, we uncovered in Lemma~\ref{th:psd} that the general form of this PSD is impulsive and thus only meaningful when integrated. Conveniently, from Lemma~\ref{th:channel_response_random} we notice that at any fixed pair $(r_z,s_z)$, the random variable $W_{\pm\pm}(k_x,k_y,\kappa_x,\kappa_y) e^{\mp\imagunit  \gamma(\kappa_x,\kappa_y) s_z} e^{\pm\imagunit  \gamma(k_x,k_y) r_z}$ is statistically equivalent to $W_{\pm\pm}(k_x,k_y,\kappa_x,\kappa_y)$ for all $(k_x,k_y,\kappa_x,\kappa_y)$. Hence, we may rewrite each one of the four channel contributions in \eqref{Fourier_spectral_random} equivalently as
\begin{align} \notag
& h(\vect{r},\vect{s}) = \frac{1}{(2\pi)^2} \iiiint_{-\infty}^{\infty} dk_x dk_y d\kappa_x d\kappa_y \,  e^{\imagunit \left(k_x r_x + k_y r_y \right)}   \\& \hspace{0.5cm} \label{Fourier_spectral_continous_channel}  e^{-\imagunit \left(\kappa_x s_x + \kappa_y s_y \right)} \sqrt{S_{\pm \pm}(k_x,k_y,\kappa_x,\kappa_y)} W_{\pm\pm}(k_x,k_y,\kappa_x,\kappa_y)
\end{align}
where equality must be understood in a distribution sense and
\begin{align} \notag
S_{\pm \pm}(k_x,k_y,\kappa_x,\kappa_y) & = \frac{A^2_{\pm \pm}(k_x,k_y,\kappa_x,\kappa_y)}{\gamma(k_x,k_y) \gamma(\kappa_x,\kappa_y)} \\& \label{psd_spectral} \hspace{1cm}  \mathbbm{1}_{\mathcal{D}\times\mathcal{D}}(k_x,k_y,\kappa_x,\kappa_y)
\end{align}
where we have embedded the integration domain into a functional dependence through an indicator function.  
We have thus shown that the channel response $h(\vect{r},\vect{s})$ between every pair of parallel $z$-oriented planes at source and receiver is exactly described by a 4D \emph{Fourier spectral representation} with parameters $r_z$ and $s_z$.
Comparing the two spectral representations in \eqref{Fourier_spectral_continous} and \eqref{Fourier_spectral_continous_channel}, we observe that the complex oscillations $e^{\imagunit \omega t}$ in \eqref{Fourier_spectral_continous} are replaced by two 2D Fourier harmonics $e^{\imagunit \left(k_x r_x + k_y r_y \right)}$ and $e^{-\imagunit \left(\kappa_x s_x + \kappa_y s_y \right)}$. Also, $S(\omega)$ is substituted by $S_{\pm \pm}(k_x,k_y,\kappa_x,\kappa_y)$ in \eqref{psd_spectral}. Summing together all possible four contributions yields the total PSD in \eqref{power},
\begin{align} \notag
&S(k_x,k_y,\kappa_x,\kappa_y) \!=\! S_{++}(k_x,k_y,\kappa_x,\kappa_y) \!+\! S_{+-}(k_x,k_y,\kappa_x,\kappa_y) \\\label{psd_sum_four} & \hspace{1cm} + S_{-+}(k_x,k_y,\kappa_x,\kappa_y) + S_{--}(k_x,k_y,\kappa_x,\kappa_y)
\end{align}
with $S_{\pm\pm}(k_x,k_y,\kappa_x,\kappa_y)$ given by \eqref{psd_spectral}.
In summary, the entire effect due to wave propagation along the arbitrarily chosen $z$-axis is purely deterministic and known a priori. A remarkable consequence of this observation is that a 3D volumetric array offers no extra DoF over a 2D planar array \cite{PizzoSPAWC20,PizzoTSP21}. The same conclusion is drawn in \cite{FranceschettiLandau} where it is pointed out that the world has only an apparent 3D informational structure, which is subject to a 2D representation. 


\subsection{Fourier Plane-Wave Series Expansion} \label{sec:non_asymptotic}

\begin{figure*}
\begin{align} \notag
P & = \frac{1}{(2\pi)^4} \iiiint_{\mathcal{D}\times\mathcal{D}} dk_x dk_y d\kappa_x d\kappa_y  \, \frac{1}{\gamma(k_x,k_y) \gamma(\kappa_x,\kappa_y)} \Big(A^2_{++}(k_x,k_y,\kappa_x,\kappa_y)   \\ \tag{68}\label{power_sum} &  \hspace{3cm}
+ A^2_{+-}(k_x,k_y,\kappa_x,\kappa_y)  + A^2_{-+}(k_x,k_y,\kappa_x,\kappa_y) + A^2_{--}(k_x,k_y,\kappa_x,\kappa_y) \Big)
\end{align}
\hrule
\begin{align} \notag
P & =  \iiiint_{\mathcal{S}_+\times\mathcal{S}_+} \!\!\!\!\!\!\!\!\!d\Omega_{\rm r} d\Omega_{\rm t} \, A^2_{++}(\theta_{\rm r},\phi_{\rm r},\theta_{\rm t},\phi_{\rm t}) + \iiiint_{\mathcal{S}_+\times\mathcal{S}_-} \!\!\!\!\!\!\!\!\!d\Omega_{\rm r} d\Omega_{\rm t}  \, A^2_{+-}(\theta_{\rm r},\phi_{\rm r},\theta_{\rm t},\phi_{\rm t})   \\ \tag{70}\label{power_angular} &  \hspace{3cm}
+  \iiiint_{\mathcal{S}_-\times\mathcal{S}_+} \!\!\!\!\!\!\!\!\!d\Omega_{\rm r} d\Omega_{\rm t}  \, A^2_{-+}(\theta_{\rm r},\phi_{\rm r},\theta_{\rm t},\phi_{\rm t})  +  \iiiint_{\mathcal{S}_-\times\mathcal{S}_-} \!\!\!\!\!\!\!\!\!d\Omega_{\rm r} d\Omega_{\rm t}  \,  A^2_{--}(\theta_{\rm r},\phi_{\rm r},\theta_{\rm t},\phi_{\rm t})
\end{align}
\hrule
\end{figure*}

Let $h(t)$ be now a bandlimited stationary Gaussian process of bandwidth $\Omega$ that is observed over an interval of duration $T$. The Karhunen-Loeve expansion of $h(t)$ provides an orthonormal description of $h(t)$ over some basis set of functions with a finite number of statistically independent coefficients \cite[Sec.~6.4]{FranceschettiBook}.
The ACF $c(t)$ is expressed by a Hilbert-Schmidt decomposition over the same basis set of functions \cite[Sec.~3.4]{FranceschettiBook}.  
However, finding this basis set is hard in practice, as an explicit solution is only available for a few cases. As an example, for a constant $S(\omega)$, it can be found by solving the Slepian's concentration problem \cite[Sec.~2]{FranceschettiBook}.
Fortunately, as the time-bandwidth product grows large, but finite, i.e., $\Omega T \gg 1$, the Karhunen-Loeve decomposition becomes a Fourier series expansion \cite[Sec.~]{VanTreesBook}.
Precisely, the Karhunen-Loeve eigenfunctions become Fourier harmonics and the associated eigenvalues statistically independent Gaussian coefficients, whose variances are obtained by sampling $S(\omega)$ at integer multiples of the fundamental frequency $2\pi/T$. This Fourier series expansion tends to the Fourier representation in \eqref{Fourier_spectral_continous}, asymptotically as $\Omega T\to \infty$.
Simply put, the Fourier spectral representation in \eqref{Fourier_spectral_continous} accomplishes the same result as $\Omega T \to \infty$ of the Karhunen-Loeve decomposition for finite $\Omega T$ values. 

The key in providing a generalization of the time-domain theory to spatial electromagnetic channels is the bandlimited property in~\eqref{psd_spectral} that naturally arise from physics considerations. Without loss of generality, we observe $h(\vect{r},\vect{s})$ over a $z$-oriented planar region of maximum dimension $L$~m.
In analogy with the time-domain case, any Karhunen-Loeve expansion of $h(\vect{r},\vect{s})$ becomes a Fourier (plane-wave) series expansion as the space-bandwidth product along each dimension grows large, but finite, i.e., $L/\lambda \to \infty$ \cite[Sec.~3]{PizzoTWC21}. The main difference with the time-domain case lies in the computation of the variances of Fourier coefficients. These cannot be obtained by sampling $S_{\pm \pm}(k_x,k_y,\kappa_x,\kappa_y)$ in~\eqref{psd_spectral} at integer multiples of the fundamental frequency $2\pi/L$ because of the singularity appearing at its denominator. Instead, we rather integrate \eqref{psd_spectral} over a neighborhood of these frequencies,  which is always possible since the PSD is singularly-integrable \cite[Sec.~3]{PizzoTWC21}, \cite[Sec.~5]{PizzoJSAC20}. 
Convergence to the Fourier spectral representation in \eqref{Fourier_spectral_continous_channel} occurs asymptotically as $L/\lambda\to \infty$.

The validity of the Fourier plane-wave series expansion is as tight as the assumption $L/\lambda \gg 1$. To show how large this value must be in order to obtain a good approximation of the Karhunen-Loeve expansion, in Fig.~\ref{fig:acf} we illustrate the autocorrelation function at receiver $c(\vect{r})$ under isotropic propagation, as there is an explicit closed-form solution in this case, namely the Clarke's formula $c(\vect{r}) = {\rm sinc}(2 r/\lambda)$ with $r = \|\vect{r}\|$. This is compared to the ACF obtained by averaging realizations of the channel response $h(\vect{r})$, each created by the Fourier plane-wave series expansion, for different $L/\lambda$ values. 
With $L/\lambda=4$, the level curves of the approximated ACF are slightly blurred, but become quite similar to those obtained with Clarke's model already for $L/\lambda=10$.

\subsection{Convergence of Fourier Spectral Representation} \label{sec:convergence}

Integral representations are subjected to a convergence criteria. 
For a stationary random process with finite average power, mean-squared-error convergence is guaranteed by Mercer's theorem \cite{VanTreesBook,FranceschettiBook}. 
Hence, due to the space-time duality leveraged above, convergence of \eqref{Fourier_spectral_continous_channel} to the actual random field in the mean-squared-error sense is guaranteed for any channel with finite average power~\eqref{power}. We expand \eqref{power} by plugging \eqref{psd_sum_four} with \eqref{psd_spectral}, which yields the final power expression in \eqref{power_sum} for some non-negative functions $A_{\pm\pm}(k_x,k_y,\kappa_x,\kappa_y)$ modeling field directionality. The convergence criteria is summarized as follows.

\begin{lemma}
The Fourier plane-wave spectral representation in~\eqref{Fourier_spectral_continous_channel} converges in the mean-squared-error sense to the actual channel impulse response for any bounded piecewise-continuous functions $A_{\pm\pm}(k_x,k_y,\kappa_x,\kappa_y)$.
\end{lemma}
\begin{proof}
Due to the boundedness assumption, we can always find a value $A<\infty$ such that $A_{\pm\pm}(k_x,k_y,\kappa_x,\kappa_y) < A$ for all $(k_x,k_y,\kappa_x,\kappa_y)$. Plugged into \eqref{power_sum} this inequality yields 
\setcounter{equation}{68} 
\begin{equation}  \label{mean_square_channel} 
P  \le  4 \left(\frac{A}{(2 \pi)^2}\iint_{\mathcal{D}}  \frac{d\kappa_x d\kappa_y}{\gamma(\kappa_x,\kappa_y)}\right)^2.
\end{equation} 
Since $\iint_{\mathcal{D}}  {1}/{\gamma(k_x,k_y)}  dk_x dk_y = \pi^2 \kappa$, then $P \le A^2 \kappa^2/4 < \infty$. Convergence follows from Mercer's theorem \cite{VanTreesBook,FranceschettiBook}.
\end{proof}
 
Physically, the above condition implies that transfer of power between every transmit and receive propagation directions is bounded across the entire angular domain. This is always satisfied in real-world propagation environments.


\section{Physical modeling of the spectral factor} \label{sec:phy_model_spectral}

While maintaining a high level of abstraction, we now show how to analytically model field directionality to fit a realistic propagation environment.
Since the average channel power is constant, due to the stationarity assumption, the small-scale fading $h(\vect{r},\vect{s})$ can thus be normalized such that it has unit average power, i.e., $P=1$. 
The PSD $S(k_x,k_y,\kappa_x,\kappa_y)$ in \eqref{power} can thus be regarded as a {power distribution function} of the channel, as it yields one when integrated. 
Conveniently, we change domain of representation in \eqref{power_sum} from wavenumber coordinates to elevation and azimuth angles $(\theta,\phi) \in [0,\pi] \times [0,2\pi)$ at source and receiver; see \eqref{wavenumber_spherical_2} and \eqref{wavenumber_spherical_rx_2}. This yields \eqref{power_angular} for some non-negative functionals $A_{\pm\pm}(\theta_{\rm r},\phi_{\rm r},\theta_{\rm t},\phi_{\rm t})$, each one including all proportionality constants due to power normalization. In \eqref{power_angular}, $\mathcal{S}_\pm$ are the upper (lower) hemisphere of unit radius (see Fig.~\ref{fig:spectral_distribution}), whereas $d\Omega = \sin \theta d\theta d\phi$ is the differential element of solid angles pointed by the direction $(\theta,\phi)$. 
Notice the disappearance of the $\gamma(\cdot,\cdot)$ terms in \eqref{power_sum}, as they are embedded into the Jacobian of the map, given by
\begin{align} \notag
\left|\frac{\partial( \kappa_x, \kappa_y)}{\partial(\theta,\phi)} \right|  & = 
\kappa^2 \cos\theta \sin\theta  \\& \label{Jacobian}  = \kappa \gamma({\kappa}_x,{\kappa}_y) \sin\theta
\end{align}
where we used $\cos\theta = \gamma({\kappa}_x,{\kappa}_y)/ \kappa$. Hence, each function $A^2_{\pm\pm}(\theta_{\rm r},\phi_{\rm r},\theta_{\rm t},\phi_{\rm t})$ becomes an angular power distribution function (PDF), say $p_{\pm\pm}(\theta_{\rm r},\phi_{\rm r},\theta_{\rm t},\phi_{\rm t})$, of the channel.

Henceforth, directionality of the field is specified by the arbitrary functions $p_{\pm\pm}(\theta_{\rm r},\phi_{\rm r},\theta_{\rm t},\phi_{\rm t})$, which uniquely parametrized the channel in the angular domain. We stress that these are the only parameters that need to be specified in our model. Clearly, an accurate fit of reality is conditioned on the availability of reliable channel measurements. Although necessary for conducting real-world research, our model is customizable to every possible propagation conditions. We next provide general guidelines on how to choose these PDF for a practical setting. Since they can be modeled separately, we focus on only one of the four contribution, e.g., the upgoing-upgoing contribution $p(\theta_{\rm r},\phi_{\rm r},\theta_{\rm t},\phi_{\rm t})$ for $(\theta_{\rm r},\phi_{\rm r},\theta_{\rm t},\phi_{\rm t}) \in \mathcal{S}_+ \times \mathcal{S}_+$ where we omit the $++$ subscript.

\subsection{Isotropic Propagation} \label{sec:isotropic}

The simplest model occurs when the angular power transfer between source and receiver is distributed uniformly on the entire angular spectrum; that is, an \emph{isotropic channel} \cite{MolischBook}. Under this setting, the scattering naturally decouples and $p(\theta_{\rm r},\phi_{\rm r},\theta_{\rm t},\phi_{\rm t})$ has a separable uniform structure at both link ends. This model has been studied in \cite[Sec.~4]{PizzoJSAC20}, at the receiver only, and leads to a constant bounded angular PDF $p(\theta_{\rm r},\phi_{\rm r})$ for all $(\theta_{\rm r},\phi_{\rm r})$. Nevertheless, this result can also be extended to the source side due to model separability,
\begin{equation} \label{isotropic_joint}
p(\theta_{\rm r},\phi_{\rm r},\theta_{\rm t},\phi_{\rm t}) = \frac{1}{(2\pi)^2}
\end{equation} 
for all $(\theta_{\rm r},\phi_{\rm r},\theta_{\rm t},\phi_{\rm t})\in \mathcal{S}_+ \times \mathcal{S}_+$ with $2\pi$ being the solid angle subtended by $\mathcal{S}_+$.

\subsection{Non-Isotropic Propagation}
 
A more realistic model that capture the angular selectivity of the scattering involves a non uniform angular PDF.
For simplicity, we assume a separable structure at source and receiver that leads to the Kronecker model \cite{MolischBook}
\begin{equation} \label{separable_spectral_factor}
p(\theta_{\rm r},\phi_{\rm r},\theta_{\rm t},\phi_{\rm t}) = p_{\rm r}(\theta_{\rm r},\phi_{\rm r}) p_{\rm t}(\theta_{\rm t},\phi_{\rm t})
\end{equation}
for some arbitrary smooth functions $p_{\rm r}(\theta_{\rm r},\phi_{\rm r})$ and $p_{\rm t}(\theta_{\rm t},\phi_{\rm t})$. 
The use of the separability assumption is confined to this Section. In general, wave propagation is coupled at both link ends, as correctly shown by a joint PDF $p(\theta_{\rm r},\phi_{\rm r},\theta_{\rm t},\phi_{\rm t})$. We focus on of the two functions in \eqref{separable_spectral_factor} indistinctly, say $p(\theta,\phi)$, as the other follows similarly.

Radio wave propagation is typically \emph{clustered} around some $N_{\rm c}$ modal propagation directions \cite[Ch.~7]{MolischBook}. Thus, $p(\theta,\phi)$ is modeled as a linear combination of all cluster contributions $p_{i}(\theta,\phi)$ with $i=1, \ldots, N_{\rm c}$, that is, the PDFs mixture
\begin{equation} \label{mixture_pdf}
p(\theta,\phi) = \sum_{i=1}^{N_{\rm c}}  w_i \, p_{i}(\theta,\phi)
\end{equation}
with non-negative weights such that $\sum_{i=1}^{N_{\rm c}} w_i=1$, which specify how the channel power is distributed in each cluster. The simplest model of each $p_{i}(\theta,\phi)$ is a uniform function over a certain angular region $\Theta_{i} \subset \mathcal{S}_{+}$ \cite{PoonDoF},
\begin{equation} \label{piecewise_constant}
p_i(\theta,\phi) = \mathbbm{1}_{\Theta_{i}}(\theta,\phi).
\end{equation} 
Despite incorporating directionality of the channel, this model is not realistic as it models discontinuous energy transitions in the angular domain -- not available in classical physics. Also, the sharp decay in the Fourier domain leads to long tails in the spatial domain with possible convergence issues.

Analytical modeling of $p_i(\theta,\phi)$ is a trade-off between mathematical tractability and model accuracy. 
In directional statistics, this issue is addressed by specifying a family of PDF that are indexed by some physical parameters, rather than a single function \cite{MardiaBook}. Desirably, we would like to have a low number of parameters that incorporates sufficient information to provide an acceptable accuracy. The perfect example is the Gaussian distribution, which is fully described by a mode $\mu$ and a standard deviation $\sigma \ge 0$, indicating the value around which the distribution is most concentrated and how spread is the distribution.
Its analogue on a 3D sphere is the von Mises-Fisher (vMF) distribution,\footnote{The classical vMF distribution is defined on a unit sphere \cite{MardiaBook}. Instead, we consider only the (upper) hemisphere and divide the PDF by a factor of $2$. This leads to no mistake as power in practical scenarios spans a narrow angular interval in elevation angle \cite{Molisch2002}.} given by \cite[Eq.~(9.3.4)]{MardiaBook}
\begin{align} 
p_{i}(\theta,\phi) & = c(\alpha_i)  e^{\alpha_i \hat{\boldsymbol{\mu}}_i^{\Ttran} \hat \krx} \\ \label{VMF_pdf}
& = c(\alpha_i) e^{\alpha_i (\sin\theta \sin\mu_{\theta,i} \cos(\phi-\mu_{\phi,i}) + \cos\theta \cos\mu_{\theta,i})}
\end{align}
which is specified by a modal direction $\hat{\boldsymbol{\mu}}_{i} \in \mathcal{S}_+$,
\begin{equation} \label{modal_direction}
\hat{\boldsymbol{\mu}}_{i} = \hat{\vect{x}} \sin \mu_{\theta} \cos \mu_{\phi}   + \hat{\vect{y}} \sin\mu_{\theta}\sin\mu_{\phi}  + \hat{\vect{z}} \cos\mu_{\theta}
\end{equation} 
with elevation and azimuth angles $(\mu_{\theta},\mu_{\phi})$ and a concentration parameter $\alpha_i \ge 0$. The former specifies the propagation direction around which the channel power is most concentrated while the latter determines the power concentration angularly; it can be regarded as an inverse standard deviation. 
Analytically, \eqref{VMF_pdf} is obtained by restricting a 3D Gaussian with circular level curves onto a unit sphere and  renormalizing by a constant $c(\alpha_i) = \alpha_i/(4\pi \sinh \alpha_i)$ to obtain a PDF \cite{Mardia1975}.\footnote{Integral over spherical supports must include the Jacobian term $\sin(\theta)$.}
More general PDFs with elliptical level curves may be considered likewise \cite{Kent}.
As limiting cases, when $\alpha_i=\infty$, then \eqref{VMF_pdf} becomes an impulsive function
\begin{equation} \label{dirac_delta}
p_i(\theta,\phi) = \delta(\hat \krx - \hat{ \boldsymbol{\mu}}_i)
\end{equation}
that implies power concentrated in only one direction $\hat{\boldsymbol{\mu}}_i$. Instead, when $\alpha_i = 0$, \eqref{VMF_pdf} reduces to
\begin{equation} \label{isotropic}
p_i(\theta,\phi) = \frac{1}{4\pi}
\end{equation}
all $(\theta,\phi)\in \mathcal{S}_+$. Multiplying \eqref{isotropic} by $2$, to account for the lower hemispherical support, and squaring the obtained result, to include the other receiver/source side, we obtain the isotropic angular PDF in \eqref{isotropic_joint}.
 
\subsection{Numerical Generation of vMF Distribution}


 \begin{figure}[t!]
\begin{subfigure}{\columnwidth}
  \includegraphics[width=.9\linewidth]{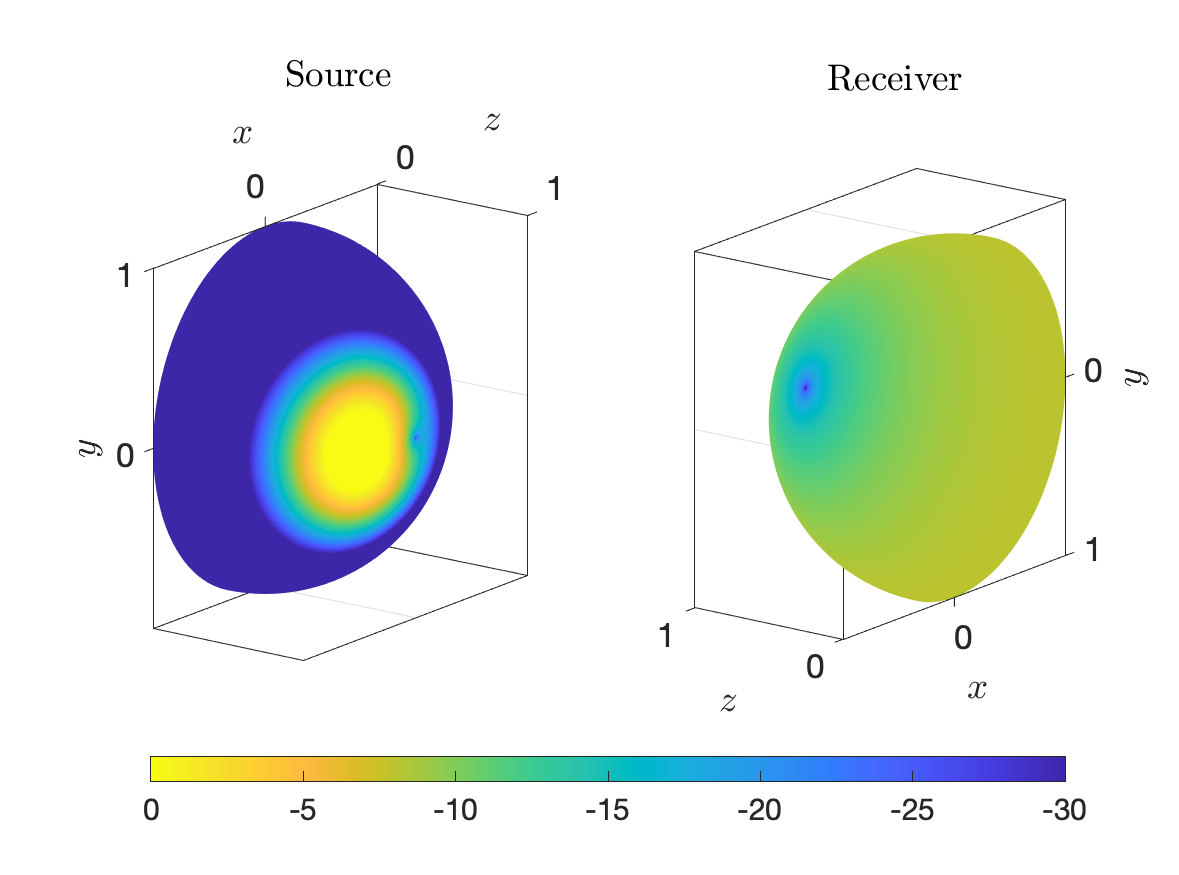}
  \caption{Single cluster (source) and isotropic propagation (receiver).}\label{fig:vMF_first}
\end{subfigure}
\begin{subfigure}{\columnwidth}
  \includegraphics[width=.9\linewidth]{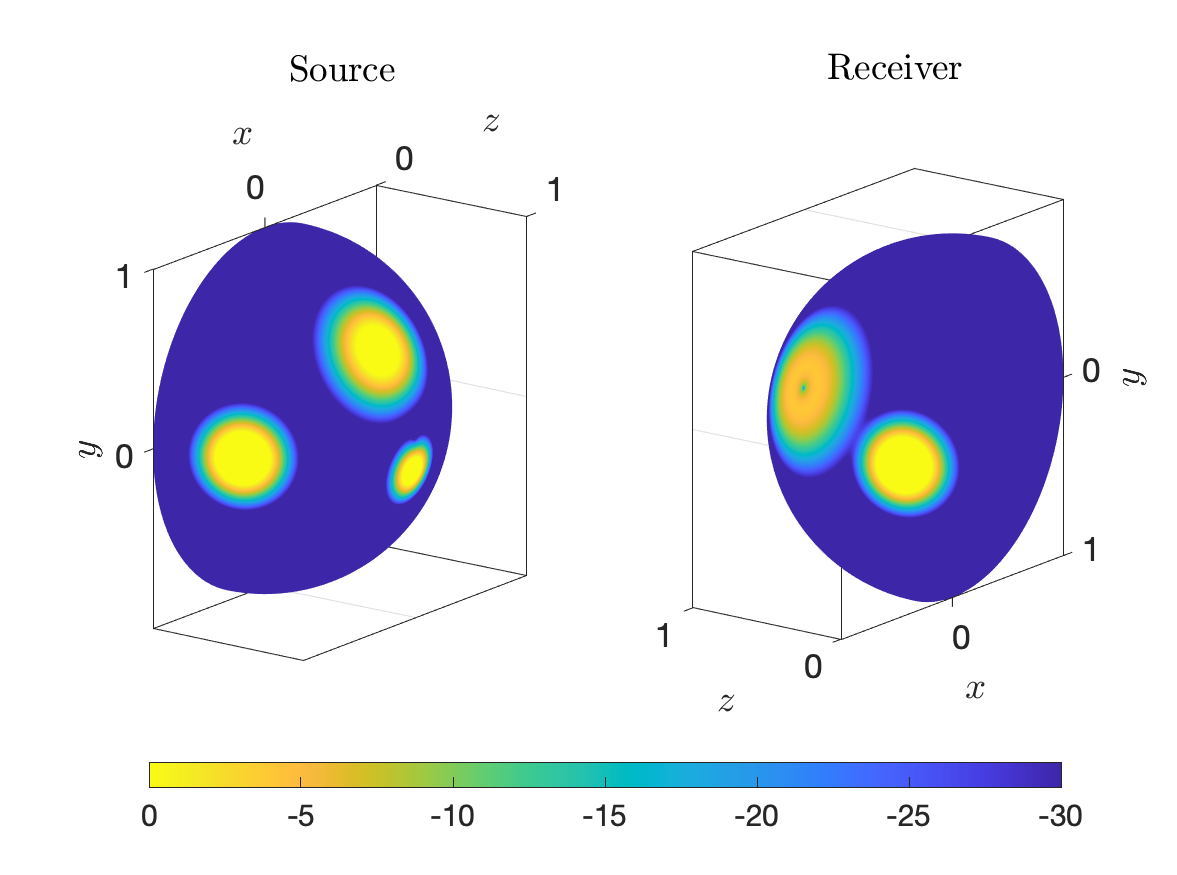}
  \caption{Multiple clusters (source and receiver).}\label{fig:vMF_second}
\end{subfigure}
\caption{Illustration of angular PDF $p(\theta_r,\phi_r,\theta_s,\phi_s)$ generated by a 3D vMF distribution with different parameters.}
\label{fig:vMF}
\end{figure}  

%
%

The vMF distribution is specified by its mode $\boldsymbol{\mu}_i$ and (normalized) variance $\nu_i^2 \in [0,1]$ for each cluster $i = 1,\ldots, N_{\rm c}$. The former is given by \eqref{modal_direction} and specified by a pair of elevation and azimuth angles $(\mu_{\theta,i},\mu_{\phi,i})$. The latter is expressed as a function of the  concentration parameter $\alpha_i$ and is obtained by solving the fixed-point equation
\begin{equation} \label{fixed_point}
\nu_i^2 = 1 - (\coth \alpha_i - 1/\alpha_i)^2
\end{equation} 
for any $\nu_i^2 \in(0,1)$. The limit case $\nu_i^2=1$ corresponding to the isotropic scenario must be treated separately by setting $p(\theta,\phi)=1/(2\pi)$ as specified in~\eqref{isotropic_joint}.
To generate the desired vMF distribution with parameters $\{w_i\}$, $N_{\rm c}$, $\{(\mu_{\theta,i},\mu_{\phi,i})\}$, and $\{\nu_i^2\}$ one should follow the following steps. First, compute the concentration parameter $\alpha_i$ by solving \eqref{fixed_point}. Second, generate the PDF in \eqref{VMF_pdf} with parameters $(\alpha_i,\mu_{\theta,i},\mu_{\phi,i})$. Third, repeat the previous steps for all clusters $i = 1,\ldots, N_{\rm c}$ and use all generated PDFs in the weighted sum in \eqref{mixture_pdf} with parameters $w_i$.  

As an example, we plot the vMF distributions $p(\theta,\phi) \sin\theta$ (inclusive of Jacobian) generated by the described method with uniform weights as a function of $(\theta,\phi)\in \mathcal{S}_+$ at source and receiver. 
In Fig.~\ref{fig:vMF_first}, we have a single cluster at the source side with $(\mu_{\theta,1},\mu_{\phi,1}) = (20^\circ,90^\circ)$ and $\nu^2_1 = 0.05$, whereas the receiver sees an isotropic propagation with $\nu^2_1 = 1$.  
In Fig.~\ref{fig:vMF_second}, multiple clustering is shown at both link ends. At the source, we have three clusters with $(\mu_{\theta,1},\mu_{\phi,1}) = (60^\circ,90^\circ)$, $(\mu_{\theta,2},\mu_{\phi,2}) = (30^\circ,15^\circ)$, and $(\mu_{\theta,3},\mu_{\phi,3}) = (10^\circ,180^\circ)$ to which correspond the variances $\nu^2_1 = 0.01$, $\nu^2_2 = 0.02$, and $\nu^2_3 = 0.005$. At the receiver, we have two clusters with $(\mu_{\theta,1},\mu_{\phi,1}) = (0^\circ,45^\circ)$ and $(\mu_{\theta,2},\mu_{\phi,2}) = (0^\circ,45^\circ)$ to which correspond $\nu^2_1 = 0.03$ and $\nu^2_2 = 0.01$.

\section{Conclusions} \label{sec:conclusions}
The continuous-space electromagnetic channel can always be modeled as an LSV system. Hence, it is fully described by its deterministic channel impulse response that can be exactly described in terms of plane waves.
This argument is supported by the availability of a closed-form Fourier plane-wave representation, which builds upon first principles of wave propagation theory.
Our formulation is the most general as it encompasses arbitrary propagation environments, abstracts from the particular array topology, and is valid irrespective of the communication range (i.e., embeds wavefront curvature even in the reactive near-field region).

When the desirable properties of Gaussian distribution and spatial stationarity are retained, a convenient statistical Rayleigh fading model is obtained in the radiative near-field region. The latter is subjected to the exclusion of reactive propagation mechanisms that unveils the low-pass spatial filtering behavior of electromagnetic channels, due to the absence of high communication modes associated with evanescent waves \cite{PizzoTSP21}. 
As for time-domain stationary Gaussian random processes \cite{VanTreesBook}, we derived a Fourier spectral representation that provides a second-order characterization of the channel in terms of statistically independent Gaussian random coefficients. This enjoys asymptotic convergence properties in the mean-squared-error sense, as as the array size becomes large compared to the wavelength \cite{PizzoJSAC20,PizzoTWC21}.

Real-world measurements are needed to correctly extracting model parameters for a prescribed environmental class.
To bring out the key concept, this paper considered scalar electromagnetic channels that physically correspond to acoustic propagation \cite{MarzettaIT}.
The electromagnetic MIMO channels is obtained by sampling the continuous-space model at antenna locations~\cite{PizzoJSAC20,PizzoTWC21}. At current stage, non-idealized antennas, mutual coupling among antenna elements \cite{Nossek}, and wideband transmissions are still missing.
Polarization effects may be incorporated by replacing every antenna point by three mutually perpendicular electric dipoles. Consequently, the vector-valued electromagnetic channel is given by a Fourier plane-wave representation encompassing horizontally polarized and vertically polarized plane waves \cite{MarzettaNokia,MarzettaVector}.


\appendices
 
\section{Proof of Theorem~\ref{th:4D_plane_wave_representation_complete}} \label{app:channel_response}

Plugging~\eqref{source_vector} with \eqref{incident_spectrum} into \eqref{receive_spectrum_vector},
 \begin{align} \label{receive_spectrum_vector_aux} 
\vect{E}_{\rm r}(k_x,k_y) & \!=\! \frac{\kappa \eta}{2}  \iint_{-\infty}^\infty \!\!\!\!\!\! d\kappa_x d\kappa_y \, \frac{\vect{K}(k_x,k_y,\kappa_x,\kappa_y)}{\gamma(\kappa_x,\kappa_y)}
\begin{bmatrix}
J_+(\kappa_x,\kappa_y) \\
J_-(\kappa_x,\kappa_y)
\end{bmatrix}
\end{align}
for all $(k_x,k_y)\in\Real^2$. Then, using \eqref{receive_spectrum_vector_aux} into the receive field expression in \eqref{received_field_vector} while replacing the source spectrum with its Fourier plane-wave transform in \eqref{source_spectrum} yields the convolutional model in \eqref{channel_NLoS}. The channel response is given by \eqref{channel_response_complete} with
\begin{equation} \label{angular_response_complete_aux}
\vect{H}(k_x,k_y,\kappa_x,\kappa_y) =  \frac{\kappa \eta}{2} \frac{\vect{K}(k_x,k_y,\kappa_x,\kappa_y)}{\gamma(\kappa_x,\kappa_y)}
\end{equation}
while $\vect{K}(\cdot,\cdot)$ is given by \eqref{propagation_kernel}.
To obtain a symmetric expression of the angular response, without loss of generality, we pull out from the propagation kernel the term $\gamma^{1/2}(\kappa_x,\kappa_y)/\gamma^{1/2}(k_x,k_y)$ so that \eqref{angular_response_complete_aux} becomes
\begin{equation} \label{angular_response_aux}
\vect{H}(k_x,k_y,\kappa_x,\kappa_y) =  \frac{\kappa \eta}{2} \frac{\vect{K}(k_x,k_y,\kappa_x,\kappa_y)}{\gamma^{1/2}(k_x,k_y) \gamma^{1/2}(\kappa_x,\kappa_y)}
\end{equation}
while $\gamma^{1/2}(\cdot,\cdot)$ is derived from \eqref{kappa_z} as
\begin{equation} \label{kappaz_sqrt}
\gamma^{1/2}(\kappa_x,\kappa_y) = 
\begin{cases}
(\kappa^2 - \kappa_x^2 - \kappa_y^2)^{1/4} &  \kappa_x^2 + \kappa_y^2\le \kappa^2 \\
\sqrt{\imagunit} (\kappa_x^2 + \kappa_y^2 - \kappa^2)^{1/4} &  \kappa_x^2 + \kappa_y^2> \kappa^2
\end{cases}
\end{equation}
where $\sqrt{\imagunit}$ should be understood as $e^{\imagunit \pi/4}$.

\section{Proof of Lemma~\ref{th:reciprocity}} \label{app:reciprocity}

\begin{figure} [t!]
        \centering
        	\includegraphics[width=.7\columnwidth,tics=10]{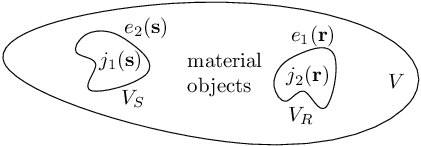}
                \caption{Geometry used for reciprocity theorem. When the source $j_1(\vect{s})$ is turned on, generates the field $e_1(\vect{r})$. When the source $j_2(\vect{r})$ is on, generates the field $e_2(\vect{s})$.} \vspace{0.0cm}
                \label{fig:reciprocity_th} 
        \end{figure}

Consider two different source distributions $j_1(\vect{s})$ and $j_2(\vect{s})$ with corresponding scalar fields $e_1(\vect{r})$ and $e_2(\vect{r})$. Each pair $(j_i(\vect{s}),e_i(\vect{r}))$ is related through the spatial convolution in \eqref{channel_NLoS} where $h_i(\vect{r},\vect{s})$ is the associated channel impulse response with $i=1,2$.
Since both fields obey the inhomogeneous Helmholtz equation in \eqref{Helm_eq}, Green's reciprocity theorem holds \cite[Eq.~(1.3.23)]{ChewBook}
\begin{equation} \label{reciprocity_relation}
\int_{V} j_1(\vect{s}) e_2(\vect{s}) \, d\vect{s} =  \int_{V} j_2(\vect{r}) e_1(\vect{r}) \, d\vect{r}
\end{equation}
for any arbitrary volume $V$ where measurements and injection operations are carried out.
Plugging \eqref{channel_NLoS} into \eqref{reciprocity_relation},
\begin{align} \notag
& \int_{V} j_1(\vect{s}) \left(\int_{\Real^3} j_2(\vect{r}) h_2(\vect{s},\vect{r}) \, d\vect{r} \right) \, d\vect{s} = \\& \label{reciprocity_relation_aux} \hspace{2cm} \int_{V} j_2(\vect{r}) \left(\int_{\Real^3} j_1(\vect{s}) h_1(\vect{r},\vect{s}) \, d\vect{s} \right) \, d\vect{r}.
\end{align}
Without loss of generality, we let $j_1(\vect{s})$ be non-zero within a volume $V_S$, whereas $j_2(\vect{r})$ is defined within another volume $V_R$. In turn, \eqref{reciprocity_relation_aux} becomes
\begin{align} \notag
& \int_{V_S} j_1(\vect{s}) \left(\int_{V_R} j_2(\vect{r}) h(\vect{s},\vect{r}) \, d\vect{r} \right) \, d\vect{s} = \\& \label{reciprocity_relation_aux_2} \hspace{2cm} \int_{V_R} j_2(\vect{r}) \left(\int_{V_S} j_1(\vect{s}) h(\vect{r},\vect{s}) \, d\vect{s} \right) \, d\vect{r}.
\end{align}
where we used $h_1(\vect{r},\vect{s}) = h(\vect{r},\vect{s})$ and applied channel reciprocity, i.e., $h_2(\vect{s},\vect{r}) = h_1(\vect{s},\vect{r})$. The above equality implies that $h(\vect{r},\vect{s})$ must be insensitive to the interchange of source and receive locations, namely $h(\vect{r},\vect{s}) = h(\vect{s},\vect{r})$ for all points. We therefore interchange $\vect{r}$ and \vect{s} in \eqref{channel_response_complete} to determine under what conditions on $\vect{H}(k_x,k_y,\kappa_x,\kappa_y)$ the above reciprocity relation holds. 
To this end, we begin with noticing the relationships
\begin{align} 
\vect{a}^*(k_x,k_y,\vect{s}) & = \vect{a}(-k_x,-k_y,\vect{s}) \\
  \vect{a}(\kappa_x,\kappa_y,\vect{r}) & = \vect{a}^*(-\kappa_x,-\kappa_y,\vect{r})
\end{align}
which follow from \eqref{array_response} plus change of sign in the $z$ component due to upgoing(downgoing) waves becoming downgoing(upgoing) waves.
After the change of integration variables $(k_x,k_y) = (-\kappa_x,-\kappa_y)$ in \eqref{channel_response_complete}, and vice versa, we obtain
\begin{equation} \label{symmetry_frequency}
\vect{H}(k_x,k_y,\kappa_x,\kappa_y) = \vect{H}^{\Ttran}(-\kappa_x,-\kappa_y,-k_x,-k_y)
\end{equation}
for all $(k_x,k_y,\kappa_x,\kappa_y)$.

\section{Proof of Theorem~\ref{th:stationary}} \label{app:stationarity}

\subsection{Gaussian Distribution}

A set of random variables $H_1, H_2, \ldots, H_N$ have circularly-symmetric complex-Gaussian joint distribution if all their linear combinations are also Gaussian \cite[Sec.~2.6]{VanTreesBook}, namely, if the random variable
\begin{equation}
h = \sum_{n=1}^N a_n H_n
\end{equation}
is circularly-symmetric complex-Gaussian for all possible coefficients $a_1, a_2, \ldots, a_N$, i.e., $h \sim \CN(0,\sigma^2)$ with any variance $\sigma^2$. The generalization of the above definition to a double-indexed 
complex random process $h(t)$ with generating functions $a_n(t)$ and $b_m(t)$ defined within $t\in(-\infty,\infty)$ yields
\begin{equation} \label{sum_sum_time}
h(t) = \sum_{n=1}^N \sum_{m=1}^N a_n(t)  H_{n,m} b_m(t)
\end{equation}
where $H_{n,m} \sim \CN(0,\sigma^2_{n,m})$ with some variances $\sigma^2_{n,m}$. Now, by letting the number $N$ increase without bound, provided the limit of \eqref{sum_sum_time} exists, we can replace the sum with an integral 
\begin{equation} \label{integral}
h(t) = \iint_{-\infty}^\infty a(\omega,t)  H(\omega,\xi) b(\xi,t) \, d\omega d\xi
\end{equation}
which implies $H(\omega,\xi)$ must be a circularly-symmetric complex-Gaussian random process in the variables $(\omega, \xi)$ for any $a(\omega,t)$ and $b(\xi,t)$. Hence, we can generally write
\begin{equation} \label{H_omega}
H(\omega,\xi) = A(\omega,\xi) W(\omega,\xi)
\end{equation}
where $W(\omega,\xi)$ is a complex white-noise field of unit variance and $A(\omega,\xi)$ is an arbitrary non-negative functional. For example, when complex exponential generating functions are chosen, i.e., $a(\omega,t)= e^{\imagunit \omega t}$ and $b(\xi,t)= e^{-\imagunit \xi t}$, we obtain a Fourier-type relationship
\begin{equation} \label{integral}
h(t) = \iint_{-\infty}^\infty e^{\imagunit \omega t}  H(\omega,\xi) e^{-\imagunit \xi t} \, d\omega d\xi.
\end{equation}
In our setup, we add another variable to $h(t)$ and replace time with space and frequency with wavenumber to obtain $h(\vect{r},\vect{s})$ in \eqref{impulse_response_Fourier}. Study of mean-squared-error convergence of the resulting stochastic integral representation is postponed to Section~\ref{sec:convergence}.
Based on the above discussion, in order for $h(\vect{r},\vect{s})$ to be a Rayleigh fading everywhere in space, $H(k_x,k_y,\kappa_x,\kappa_y)$ in \eqref{H_spectrum} must be a circularly-symmetric complex-Gaussian random field in the variables $(k_x,k_y,\kappa_x,\kappa_y)$. 
 In turn, each entry of $\vect{H}(k_x,k_y,\kappa_x,\kappa_y)$ (and so each entry of $\vect{K}(k_x,k_y,\kappa_x,\kappa_y)$ due to \eqref{angular_response_complete}) must have joint Gaussian distribution.
A general expression for the propagation kernel is as follows
\begin{equation} \label{kernel_gaussian}
\vect{K}(k_x,k_y,\kappa_x,\kappa_y) =  \vect{A}(k_x,k_y,\kappa_x,\kappa_y) \odot \vect{W}(k_x,k_y,\kappa_x,\kappa_y)
\end{equation}
where $\vect{A}(k_x,k_y,\kappa_x,\kappa_y) \in\Real_+^{2\times 2}$ whose entries are arbitrary non-negative functions and $\vect{W}(k_x,k_y,\kappa_x,\kappa_y) \in\Complex^{2\times 2}$ is a random matrix with circularly-symmetric complex-Gaussian random field entries of unit variance, i.e.,
\begin{equation}
[\vect{W}]_{i,\ell}(k_x,k_y,\kappa_x,\kappa_y) \sim \CN(0,1)
\end{equation}
for all $(k_x,k_y,\kappa_x,\kappa_y)$ and $i,\ell = 1,2$. Notice that statistical correlation is allowed among entries of $\vect{W}(k_x,k_y,\kappa_x,\kappa_y)$ and within each entry. 

\subsection{Spatial Stationarity}

A random process $h(t)$ is said to be stationary if its joint distribution is invariant to time-shifts, i.e., $h(t)$ and $h(t-\tau)$ have the same distribution for all $\tau$. The shifted process obtained from \eqref{integral} reads as
\begin{equation} \label{integral_2}
h(t-\tau) = \iint_{-\infty}^\infty e^{\imagunit \omega t}  \left(H(\omega,\xi) e^{\imagunit (\xi-\omega)\tau}\right) e^{-\imagunit \xi t} \, d\omega d\xi.
\end{equation}
For a circularly-symmetric complex-Gaussian random process $h(t)$, stationarity is achieved by requiring $H(\omega,\xi)$ to have statistically uncorrelated entries for all $(\omega,\xi)$. Under this assumption, the joint distribution of $H(\omega,\xi)$ is determined from its marginals, each one of which is not affected by a phase shift being of circularly-symmetric complex-Gaussian distribution.  
Mapping this condition onto \eqref{H_omega} yields
\begin{equation} \label{white_noise}
\Ex\{W(\omega,\xi) W^*(\omega^\prime,\xi^\prime) \} = \delta(\omega - \omega^\prime) \delta(\xi - \xi^\prime).
\end{equation}
Altogether, $W(\omega,\xi)$ must be a white-noise complex random field of unit variance.
The weaker form of wide-sense stationarity of $h(t)$ holds as well. This may also be verified analytically by using \eqref{white_noise} into the computation of the correlation function of $h(t)$.

Generalization of the above result to spatially-stationary electromagnetic channels $h(\vect{r},\vect{s})$ requires the entries of $\vect{K}(k_x,k_y,\kappa_x,\kappa_y)$ be uncorrelated with each other and individually having correlation function proportional to a Dirac delta function in the wavenumber domain. Mapping this condition onto \eqref{kernel_gaussian} yields
\begin{align} \label{kernel_acf_tot}
& \Ex\{[\vect{W}]_{i,\ell}(k_x,k_y,\kappa_x,\kappa_y)   [\vect{W}]_{i^\prime,\ell^\prime}^*(k_x,k_y,\kappa_x,\kappa_y)\} = \delta_{i i^\prime} \delta_{\ell \ell^\prime}  \\\notag   
&   \Ex\{[\vect{W}]_{i,\ell}(k_x,k_y,\kappa_x,\kappa_y)   [\vect{W}]_{i,\ell}^*(k^\prime_x,k^\prime_y,\kappa^\prime_x,\kappa^\prime_y)\}   \\& \label{kernel_acf} \hspace{.8cm} =  \delta(k_x - k^\prime_x) \delta(k_y - k^\prime_y) \delta(\kappa_x - \kappa^\prime_x)  \delta(\kappa_y - \kappa^\prime_y).
\end{align}
Compared to the time-domain case, there is an additional requirement in the spatial domain. Notice that the generating functions in \eqref{integral} are pure oscillating complex exponentials. Hence, the power of $h(t)$ averaged over all possible realizations is invariant to any time shift. However, we recall that the Fourier plane-wave representation in \eqref{channel_response_complete} comprises two types of complex exponentials leading to propagating and evanescent waves, as discussed in Section~\ref{sec:wavenumber_response_LOS}. Since the power carried by evanescent waves decays along the $z$-axis, for spatial stationarity we must require the propagation kernel to vanish in the evanescent region. Analytically, we embed this condition into $\vect{A}(k_x,k_y,\kappa_x,\kappa_y)$ in \eqref{kernel_gaussian} by limiting its wavenumber support to a domain
\begin{equation} \label{domain}
(\kappa_x,\kappa_y,k_x,k_y)\in \mathcal{D} \times \mathcal{D}
\end{equation}
where $\mathcal{D}$ is given by \eqref{disk_T}. We notice that dispersive media having complex-valued wavenumber $\kappa$ inevitably lead to a non-stationary representation of the channel, being their effect similar to the one created by evanescent waves. 
Differently than evanescent waves, however, spatial stationarity is never achieved in this case.

\section{Proof of Lemma~\ref{th:psd}} \label{app:psd}


By equating \eqref{autocorrelation_Fourier_plane_wave} with \eqref{autocorrelation3} while plugging \eqref{2D-plane-wave} we obtain
\begin{align}  \notag 
& \boldsymbol{\phi}^{\Htran}(k_x,k_y,r_z) \vect{S}(k_x,k_y,\kappa_x,\kappa_y) \boldsymbol{\phi}(\kappa_x,\kappa_y,s_z)   = \\ \label{spectral_density_inverse1}  & \hspace{0cm} \frac{1}{(2\pi)^2}
\iint_{-\infty}^\infty S(k_x,k_y,k_z,\kappa_x,\kappa_y,\kappa_z) e^{\imagunit (k_z r_z - \kappa_z s_z)} \, dk_z d\kappa_z.
  \end{align}
Replacing $\vect{S}(k_x,k_y,\kappa_x,\kappa_y)$ with its expression in \eqref{spectral_matrix}, the left-hand side of \eqref{spectral_density_inverse1} becomes
\begin{align}  \notag
& \frac{\boldsymbol{\phi}^{\Htran}(k_x,k_y,r_z)}{\gamma(k_x,k_y)} (\vect{A}\odot \vect{A})(k_x,k_y,\kappa_x,\kappa_y)   \frac{\boldsymbol{\phi}(\kappa_x,\kappa_y,s_z)}{\gamma(k_x,k_y)} = \\ \notag
& A^2_{++}(k_x,k_y,\kappa_x,\kappa_y) e^{\imagunit  \gamma(k_x,k_y) r_z}  e^{-\imagunit  \gamma(\kappa_x,\kappa_y) s_z}   \\ \notag
& + A^2_{+-}(k_x,k_y,\kappa_x,\kappa_y) e^{-\imagunit  \gamma(\kappa_x,\kappa_y) s_z}  e^{-\imagunit  \gamma(k_x,k_y) r_z}   \\ \notag
& + A^2_{-+}(k_x,k_y,\kappa_x,\kappa_y) e^{\imagunit  \gamma(\kappa_x,\kappa_y) s_z}  e^{\imagunit  \gamma(k_x,k_y) r_z}  \\ \label{psd_total_spectral}
& + A^2_{--}(k_x,k_y,\kappa_x,\kappa_y) e^{\imagunit  \gamma(\kappa_x,\kappa_y) s_z}  e^{-\imagunit  \gamma(k_x,k_y) r_z}
  \end{align}
where $A_{\pm\pm}(\cdot,\cdot)$ are the non-negative functions included into $\vect{A}(k_x,k_y,\kappa_x,\kappa_y)$ in \eqref{angular_response_stationary}.
As an intermediate step, notice that the composition of the Dirac delta function $\delta(x)$ with a differentiable function $g(x)$ with non-zero derivative yields
\begin{equation}
\delta\big(g(x)\big) = \sum_i \frac{\delta(x - x_i)}{|\frac{\partial g(x)}{\partial x}|_{x=x_i}}
\end{equation}
for $g(x_i) = 0$ with  $i=1,2,\ldots$. For example, when $g(\kappa_z) = \kappa_z^2 - \gamma^2$ (at source) we obtain
\begin{equation} \label{delta_tx}
\delta\big(\kappa_z^2 - \gamma^2\big) = \frac{\delta(\kappa_z - \gamma) + \delta(\kappa_z + \gamma)}{2 \, \gamma}.
\end{equation}
The sampling property of the Dirac delta function yields
\begin{equation} \label{sampling_deltatx}
 \int_{-\infty}^\infty \delta\big(\kappa_z^2 - \gamma^2\big) e^{-\imagunit \kappa_z s_z} \, d\kappa_z =  \frac{e^{-\imagunit \gamma s_z} + e^{\imagunit \gamma s_z}}{2 \gamma}
\end{equation}
and (at receiver),
\begin{equation} \label{sampling_deltarx}
 \int_{-\infty}^\infty \delta\big(k_z^2 - \gamma^2\big) e^{\imagunit k_z r_z} \, dk_z =  \frac{e^{\imagunit \gamma r_z} + e^{-\imagunit \gamma r_z}}{2 \gamma}
\end{equation}
where we change the sign of the complex exponentials and replace transmit coordinates with the receive counterparts. Finally, \eqref{psd_final} is obtained by equating \eqref{spectral_density_inverse1} with \eqref{psd_total_spectral} while using \eqref{sampling_deltatx} and \eqref{sampling_deltarx}.

\bibliographystyle{IEEEbib}
\bibliography{refs_IT}

\end{document}